\documentclass[a4paper,10pt]{article}

\usepackage{amssymb}
\usepackage{amsmath}
\usepackage{subfigure}
\usepackage{slashed} 
\usepackage{marvosym}

\allowdisplaybreaks

\usepackage{soul}

\usepackage{graphicx}
\usepackage{amsthm}
\usepackage{latexsym}
\usepackage[dvips]{epsfig}

\usepackage{wasysym}
\usepackage{mathrsfs}
\usepackage{eufrak}
\usepackage{bm}
\usepackage{authblk}
\usepackage{slashed}
\usepackage{yhmath} 
\usepackage{stmaryrd}

\theoremstyle{plain}
\newtheorem{proposition}{Proposition}
\newtheorem{lemma}{Lemma}
\newtheorem{theorem}{Theorem}
\newtheorem{assumption}{Assumption}

\newtheorem*{corollary}{Corollary}
\newtheorem*{main}{Theorem}

\newtheorem{remark}{Remark}

\def\bma{{\bm a}}
\def\bmb{{\bm b}}
\def\bmc{{\bm c}}

\def\bme{{\bm e}}

\def\bmg{{\bm g}}
\def\bmh{{\bm h}}
\def\bmi{{\bm i}}
\def\bmj{{\bm j}}

\def\bmzero{{\bm 0}}
\def\bmone{{\bm 1}}

\def\bmA{{\bm A}}
\def\bmB{{\bm B}}
\def\bmC{{\bm C}}

\def\bmX{{\bm X}}

\def\bmalpha{{\bm \alpha}}

\def\bmeta{{\bm \eta}}

\def\bmmu{{\bm \mu}}

\def\bmphi{{\bm \phi}}

\def\bmsigma{{\bm \sigma}}

\def\bmpartial{{\bm \partial}}

\newcounter{mnotecount}

\newcommand{\mnotex}[1]
{\protect{\stepcounter{mnotecount}}$^{\mbox{\footnotesize $\bullet$\themnotecount}}$ 
\marginpar{
\raggedright\tiny\em
$\!\!\!\!\!\!\,\bullet$\themnotecount: #1} }

\newcommand{\bit}{\begin{itemize}}
\newcommand{\eit}{\end{itemize}}
\newcommand{\ben}{\begin{enumerate}}
\newcommand{\een}{\end{enumerate}}

\newcommand{\beq}{\begin{equation}}
\newcommand{\eeq}{\end{equation}}
\newcommand{\bea}{\begin{eqnarray}}
\newcommand{\eea}{\end{eqnarray}}

\newcommand{\bwt}{\begin{widetext}}
\newcommand{\ewt}{\end{widetext}}

\usepackage[utf8]{inputenc}
\usepackage[english]{babel}
 
\usepackage{multicol}
\usepackage{color}
 
\usepackage{comment}
 
\begin{document}

\title{\textbf{The Maxwell-scalar field system near spatial infinity }}

\author[,1]{Marica Minucci \footnote{E-mail address:{\tt
      m.minucci@qmul.ac.uk}}} 
\author[,2]{Rodrigo Panosso Macedo \footnote{E-mail address:{\tt
      r.panossomacedo@soton.ac.uk}}} 
\author[,1]{Juan A. Valiente
  Kroon \footnote{E-mail address:{\tt j.a.valiente-kroon@qmul.ac.uk}}}
\affil[1]{School of Mathematical Sciences, Queen Mary, University of
  London, Mile End Road, London E1 4NS, UK.}
\affil[2]{School of Mathematical Sciences, University of Southampton, Highfield Southampton SO17 1BJ, UK.}

\maketitle

\begin{abstract}
We make use of Friedrich's representation of spatial infinity to study
asymptotic expansions of the Maxwell-scalar field system near spatial
infinity. The main objective of this analysis is to understand the
effects of the non-linearities of this system on the regularity of
solutions and polyhomogeneous expansions at null infinity and, in particular, at the critical sets
where null infinity touches spatial infinity. The main outcome from
our analysis is that the nonlinear interaction makes both fields more
singular at the conformal boundary than what is seen when the fields are non-interacting. In
particular, we find a whole new class of logarithmic terms in the
asymptotic expansions which depend on the coupling constant between
the Maxwell and scalar fields. We analyse the implications of these
results on the peeling (or rather lack thereof) of the fields at null infinity.
\end{abstract}



\section{Introduction}
Among the main open problems in the mathematical relativity there is
the so-called \textit{problem of spatial infinity} ---see
e.g. \cite{Fri18}. This problem concerns the understanding of the
consequences arising from the degeneracy of the conformal structure of
the spacetime at spatial infinity. A systematical method to tackle
this problem goes back to the seminal work of Friedrich
\cite{Fri98a}. The key idea of this work is the development of a
representation of spatial infinity, the so-called \textit{F-gauge},
which allows the formulation of a regular Cauchy problem in a
neighbourhood of spatial infinity for the \textit{conformal Einstein
  field equations}. In this setting it is possible to show that,
unless the initial data is fine-tuned, the solutions to the conformal
Einstein field equations develop two types of logarithmic
singularities at the \textit{critical sets} $\mathcal{I}^\pm$ where null infinity meets
spatial infinity. There are logarithmic singularities associated to
the linear part of the equations and the ones associated to the
nonlinear equations which appear at higher order in the expansion.  In
the particular case of \textit{time-symmetric initial data sets} for
the Einstein field equations which admit a point compactification at
infinity for which the resulting \textit{conformal metric} is
analytic, it is possible to show that a certain subset of the
logarithmic singularities can be avoided if the conformal metric
$\bmh$ satisfies the conformally invariant condition
\[
D_{\{i_p \dots i_1} b_{jk\}}=0, \qquad p=0, 1, 2, \dots, 
\]
where $b_{jk}$ denotes the Cotton-Bach tensor of the metric $\bmh$ and
$\{\dots\}$ denotes the operation of computing the symmetric tracefree
part, in particular if $\bmh$ is conformally flat then
$b_{jk}=0$. Although this condition is necessary to avoid
logarithmic singularities at the critical sets it is not
sufficient. It has been shown that static solutions to the Einstein
field equations are logarithmic free at the critical points of
Friedrich's representation of spatial infinity. Moreover, the analysis
in \cite{Val10,Val11} strongly suggests the conjecture that, among the
class of time symmetric initial data sets, only those which are static
in a neighbourhood of infinity will give rise to developments which
are free of logarithmic singularities at the critical sets. The gluing
techniques developed in e.g. \cite{Cor00,ChrDEl02} allow the
construction of large classes of initial data sets with this property.

\medskip
In general, linearised field propagating on the Minkowski spacetime
also develop logarithmic singularities at the critical sets ---see
e.g. \cite{Val07b,Val09a}. In particular, the Maxwell 
field system provides useful insights to study the linearised
gravitational field and as a model for the Bianchi equations satisfied
by the components of the Weyl tensor. Looking beyond linear model problems for the Einstein field equations, it is natural to look for 
systems which can be used to unterstand the effects of the
non-linear interactions on the regularity of
solutions at the conformal boundary. In the present article we
consider the possibility of using the \emph{Maxwell-scalar field system on
the Minkowski spacetime} to
this purpose. More precisely, we develop a theory for the solutions
to these equations in a neighbourhood of spatial infinity ---in
particular, the \emph{solution jets} at the cylinder at spatial
infinity\footnote{Roughly speaking, a solution jet of order $p$ is the
  restriction of the solution and its radial derivatives up to order
  $p$ at the cylinder at spatial infinity. The elements of the jet of
  order $p$ can be thought of as the coefficients in a Taylor-like
  expansion. The precise definition can be found in Section \ref{Section:AsymptoticExpansions}. }. This can be
done by studying their asymptotic expansions near spatial infinity
with a technique that goes under the name of \emph{F-expansions}.
This construction exploits the fact that the cylinder at spatial
infinity, $\mathcal{I}$, is a \emph{total characteristic} of the
evolution equations associated to the Maxwell-scalar system.
Accordingly, the evolution equations reduce to an interior system
(\emph{transport equations}) upon evaluation on the cylinder
$\mathcal{I}$. These transport equations allow to relate properties of
the initial data, as defined on a fiduciary initial hypersurface
$\mathcal{S}_\star$, with radiative properties of the solution which
are defined at null infinity $\mathcal{I}^\pm$ and fully determine the
solution jets on the cylinder at spatial infinity. The main outcome of
this analysis is contained in the following theorem:

\begin{main}[Main theorem, rough version]
For generic analytic data for the Maxwell-scalar field system with finite
energy, the solution jets on the
cylinder at spatial infinity $\mathcal{I}$ develop logarithmic singularities
at the critical sets $\mathcal{I}^\pm$.
\end{main}

In other words, generic solutions to the Maxwell-scalar field system
are singular at the critical sets $\mathcal{I}^\pm$. Under the further
assumption that these singularities propagate along null infinity, it
is possible to analyse the consequences of these singularities on the
\emph{peeling} properties of the Maxwell and scalar fields. One has the
following corollary:

\begin{corollary}
If the solution jets give rise to a solution to the Maxwell-scalar
field system near $\mathcal{I}$, then the
Maxwell-scalar field system generically has logarithmic singularities which
spread along the conformal boundary destroying the smoothness of
Faraday tensor and scalar field tensor along the conformal
boundary. In particular, there is no classical peeling behaviour at
null infinity.
\end{corollary}

Although the content of our Main Theorem is analogous to what it is
obtained in the case of the Einstein field equations, \emph{the detailed
analysis leading to the result shows that, in fact, the Maxwell-scalar
system is not a good model problem as the elements of the solution
jets are more singular at the critical sets than what a direct
extrapolation from the Einstein field equations would suggest.} This
new singular behaviour can be traced back to the cubic coupling
between the Maxwell and scalar fields.The latter is the most
important insight obtained from our analysis. 

\subsection*{Outline of the paper}
This article is structured as follows: Section
\ref{Section:CylinderAtspatialInfinity} provides a brief discussion of
Friedrich's representation of spatial infinity for the Minkowski
spacetime and the coordinate and frame gauge associated to this
description.   Section \ref{Section:MSFEqns} provides a discussion of
the Maxwell-scalar field system which is geared towards the
analysis in this article. In particular, it provides a description of
the system in terms of the space-spinor formalism ---to the best of
our knowledge this approach is new in the literature. The conformal
properties of the system are also analysed. Section
\ref{StructuralProperties} provides a discussion of the structural
properties of the Maxwell-scalar field system in relation to
Friedrich's representation of spatial infinity. In particular, it is
explained how these structural properties can be used to construct
solution jets at the cylinder at spatial infinity whose elements are
completely determined by initial data for the fields at some fiduciary
Cauchy hypersurface.  Section \ref{Section:InitialConditions}
discusses the construction of initial data for the Maxwell-scalar
field system. Section \ref{Section:AsymptoticExpansions} provides a
detailed analysis of the properties of the solution jets at the
cylinder at spatial infinity. Of particular interest in the analysis
is the behaviour of the elements of the jets at the critical sets where
spatial infinity meets null infinity. For completeness and as a
reference for completeness, we also
provide a discussion of the decoupled case where the constant
$\mathfrak{q}$ which couples the Maxwell and scalar fields
vanishes. Section \ref{Section:Peeling} explores the implications of
the main analysis of the article for the peeling properties of the
field. Some brief conclusions are presented in Section
\ref{Conclusions}. In addition to the above, the article contains four technical
appendices to ease the presentation of the main text. Appendix
\ref{Appendix:RicciTensor} provides details about some of the
underlying geometric structures arising in Friedrich's representation
of spatial infinity. Appendix \ref{Appendix:JacobiODE} summarises
well-know properties of polynomial solutions to the Jacobi ordinary
differentail equation. Appendix \ref{Appendix:SeriesSolution}
discusses the construction of solutions to the Jacobi ordinary
differential equation using Frobenius's method. Finally, Appendix
\ref{Appendix:VariationParameters} provides details on the construction of solutions to
inhomogeneous Jacobi equations using the method of variation of parameters.

\subsection*{Notations and Conventions}

 The signature convention for (Lorentzian) spacetime metrics will be $
 (+,-,-,-)$.  In the rest of this article $\{_a ,_b , _c ,
 . . .\}$ denote spacetime abstract tensor indices and $\{_\bma ,_\bmb , _\bmc ,
 . . .\}$ will be used as spacetime frame indices taking the values ${
   0, . . . , 3 }$.  In this way, given a basis
$\{\bme_{\bma}\}$ a generic tensor is denoted by $T_{ab}$ while its
components in the given basis are denoted by $T_{\bma \bmb}\equiv
T_{ab}\bme_{\bma}{}^{a}\bme_{\bmb}{}^{b}$. The Greek indices
${}_\mu,\, {}_\nu,\ldots$ denote spacetime coordinate indices while
the indices ${}_\alpha,\, {}_\beta,\ldots$ denote spatial coordinate indices.

\medskip
 Part of  the analysis will require the use of spinors.
In this respect, the notation and
 conventions of Penrose \& Rindler \cite{PenRin84} will be followed.
 In particular, capital Latin indices $\{ _A , _B , _C , . . .\}$ will
 denote abstract spinor indices while boldface capital Latin indices
 $\{ _\bmA , _\bmB , _\bmC , . . .\}$ will denote frame spinorial
 indices with respect to  a specified spin dyad ${
   \{\epsilon_\bmA{}^{A} \} }.$

\medskip
The  conventions for the curvature tensors are fixed by the relation
\[
(\nabla_a \nabla_b -\nabla_b \nabla_a) v^c = R^c{}_{dab} v^d.
\]

\section{The cylinder at spatial infinity and the F-gauge}
\label{Section:CylinderAtspatialInfinity}
The purpose of this section is to provide a succinct discussion of
Friedrich's representation of the neighbourhood of spatial infinity
for the Minkowski spacetime. Further details on this construction can
be found in \cite{Fri98a,Val03a,GasVal20}. A discussion of the
relation between this representation of spatial infinity and other
representations can be found in \cite{MagVal21}. 

\subsection{Conformal extensions of the Minkowski spacetime}
We start with the Minkowski metric $\tilde{\bmeta}$ written in Cartesian
coordinates $(\tilde{x}^{\mu})= (\tilde{t},\tilde{x}^{\alpha})$,
 \[
 \tilde{\bmeta}=\eta_{\mu\nu}\mathbf{d}\tilde{x}^{\mu}\otimes\mathbf{d}\tilde{x}^{\nu},
 \]
 where $\eta_{\mu\nu}=\text{diag}(1,-1,-1,-1)$. By introducing spherical
 coordinates defined by $\tilde{\rho}^2\equiv 
 \delta_{\alpha\beta}\tilde{x}^{\alpha}\tilde{x}^{\beta}$ where
 $\delta_{\alpha\beta}=\text{diag(1,1,1)}$, and an arbitrary choice of
 coordinates on $\mathbb{S}^2$, the metric $\tilde{\bmeta}$ can be
 written as
\begin{equation*}
\tilde{\bmeta}=\mathbf{d}\tilde{t}\otimes\mathbf{d}\tilde{t}
-\mathbf{d}\tilde{\rho}\otimes \mathbf{d}\tilde{\rho}-\tilde{\rho}^2
\mathbf{\bm\sigma},
\end{equation*}
with $\tilde{t}\in(-\infty, \infty)$, $\tilde{\rho}\in [0,\infty)$ and where
  $\bm\sigma$ denotes the standard metric on $\mathbb{S}^2$.  A
  strategy to construct a conformal representation of the Minkowski
  spacetime close to $i^{0}$ is to make use of  \emph{inversion coordinates} $(x^{\alpha})=(t,x^{i})$ defined by ---see
  \cite{Ste91}---
 \[ 
x^{\mu}=-{\tilde{x}^{\mu}}/{\tilde{X}^2}, \qquad \tilde{X}^2 \equiv
 \tilde{\eta}_{\mu\nu}\tilde{x}^{\mu}\tilde{x}^{\nu},
\]
which is valid in the domain
\[
\tilde{\mathcal{D}}\equiv \{ p\in \mathbb{R}^4 \; |\;
\eta_{\mu\nu} \tilde{x}^\mu(p) \tilde{x}^\nu(p) <0 \}.
\]

\begin{figure}[t]
\centering
\includegraphics[width=0.9\textwidth]{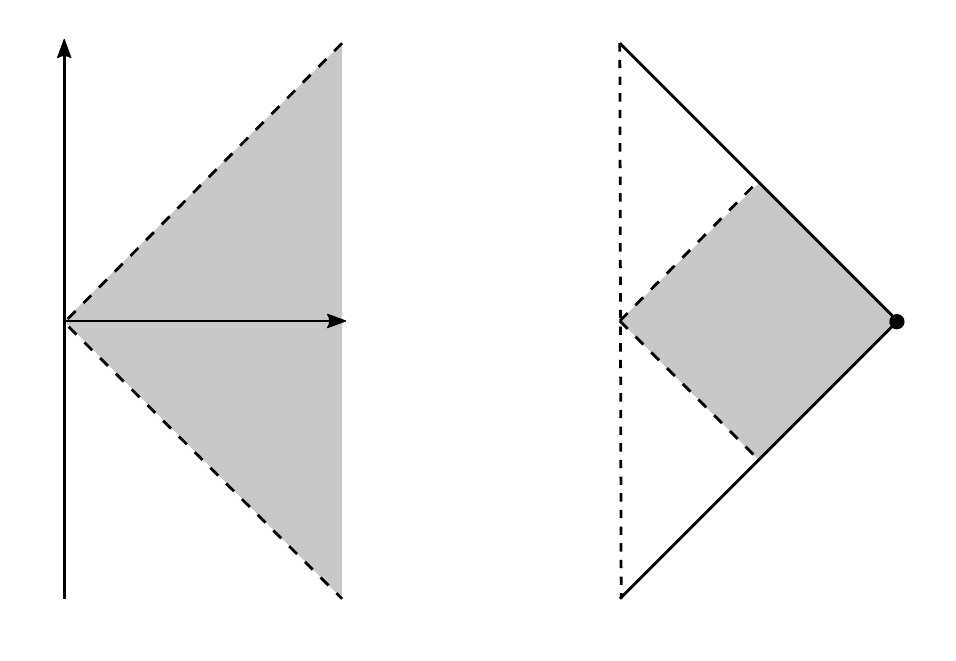}
\put(-365,240){$\tilde{t}$}
\put(-300,120){$\tilde{\mathcal{D}}$}
\put(-250,120){$\tilde{r}$}
\put(-70,140){$\mathcal{D}$}
\put(-10,130){$i^0$}
\caption{Left: The region $\tilde{\mathcal{D}}$, the complement of the
  light cone through the origin, in the physical Minkowski
  spacetime. Intuitively, this region contains spatial
  infinity. Right: the corresponding region $\mathcal{D}$ in the
  Penrose diagram of the Minkowski spacetime. }
\label{Figure:NeighbourhoodSpatialInfinity}
\end{figure}
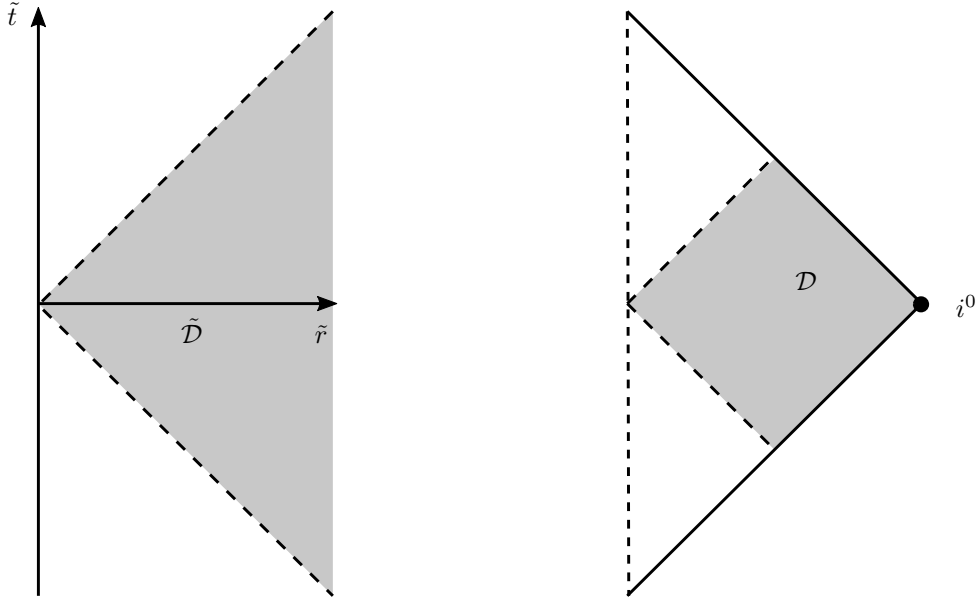

 The inverse transformation is given by
 \[
\tilde{x}^{\mu}=-x^{\mu}/X^2, \qquad
 X^2=\eta_{\mu\nu}x^{\mu}x^{\nu}.
\]
Observe, in particular that $X^2=1/\tilde{X}^2$. Using these
coordinates one identifies a conformal representation of the Minkowski
spacetime with \emph{unphysical metric} given by
\[
\bar \bmeta=\Xi^2  \tilde{\bmeta}, \qquad \Xi \equiv X^2,
\]
where 
\[
\bar\bmeta=\eta_{\mu\nu}\mathbf{d}x^{\mu}\otimes \mathbf{d}x^{\nu}.
\]
 Introducing an \emph{unphysical radial coordinate} via
the relation $\rho^2\equiv
\delta_{\alpha\beta}x^{\alpha}x^{\beta}$, one finds that the metric $\bar\bmeta$ can be
written as
\[
\bar\bmeta=\mathbf{d}t\otimes\mathbf{d}t -\mathbf{d}\rho\otimes
\mathbf{d}\rho-\rho^2 \mathbf{\bm\sigma}, \qquad \Xi=t^2-\rho^2,
\]
with $t\in(-\infty,\infty)$ and $\rho\in (0,\infty)$.
In this conformal representation, spatial infinity $i^{0}$ corresponds
to the origin of the domain
\[
\mathcal{D}\equiv \{ p\in \mathbb{R}^4 \; |\;
\eta_{\mu\nu} x^\mu(p) x^\nu(p) <0 \}.
\]
This region contains the asymptotic region of the Minkowski spacetime
around spatial infinity. Observe that $(\tilde{t},\tilde{\rho})$
are related to $(t,\rho)$ via
\[
\tilde{t}=-\dfrac{t}{t^2-\rho^2}, \qquad  \tilde{\rho}= -\dfrac{\rho}{t^2-\rho^2}.
\]
 Finally, introducing a time coordinate
$\tau$ through the relation  $t=\rho\tau$ one finds that the metric
$\bar\bmeta$ can be written as
\begin{equation*}
\bar\bmeta =\rho^2 \mathbf{d}\tau\otimes \mathbf{d}\tau
-(1-\tau^2)\mathbf{d}\rho \otimes \mathbf{d}\rho + \rho\tau
\mathbf{d}\rho\otimes \mathbf{d}\tau + \rho\tau \mathbf{d}\tau \otimes
\mathbf{d}\rho - \rho^2 \bmsigma.
\end{equation*}

\subsection{The cylinder at spatial infinity}
\label{TheCylinderAtSpatialInfinity}
The conformal representation containing the \emph{cylinder at spatial
infinity} is obtained by considering the rescaled metric
\begin{equation}
\bmeta \equiv \dfrac{1}{\rho^2} \bar\bmeta.
\label{MetricCylinder}
\end{equation}
Introducing the coordinate  $\varrho\equiv-\ln \rho$ the metric $\bmeta$
can be reexpressed as 
\[
\bmeta=\mathbf{d}\tau \otimes \mathbf{d}\tau
-(1-\tau^2)\mathbf{d}\varrho \otimes \mathbf{d}\varrho
- \tau \mathbf{d}\tau \otimes \mathbf{d}\varrho -\tau
\mathbf{d}\varrho \otimes \mathbf{d}\tau -\bm\sigma.
\]
Observe that  spatial infinity $i^{0}$, which is at infinity 
 respect to the metric $\bmeta$, corresponds to a set
which has the topology of $\mathbb{R}\times\mathbb{S}^2$ ---see
\cite{Fri98a, AceVal11}.  Following the previous discussion, 
 one considers the conformal extension $(\mathcal{M},\bmeta)$ where
\[
\bmeta=\Theta^2 \tilde{\bmeta}, \qquad \Theta\equiv \rho(1-\tau^2),
\]
and 
\[
\mathcal{M} \equiv \big\{ p \in \mathbb{R}^4 \; \rvert \; -1 \leq \tau \leq 1 , \; \; \rho(p)\geq 0\big\}.
\]
In this representation future and past null infinity are described by
the sets
\[
 \mathscr{I}^{+} \equiv \big\{ p \in \mathcal{M} \; \rvert\; \tau(p) =1 \big\}, 
\qquad \mathscr{I}^{-} \equiv \big\{ p \in \mathcal{M} \; \rvert \; 
 \tau(p) =-1\big\},
\]
while the physical Minkowski spacetime can be
identified with the set
\[
\tilde{\mathcal{M}} \equiv \big\{ p \in \mathcal{M} \; \rvert \; -1<\tau(p)<1 , \; \;\rho(p)>0 \big\}.
\]
In addition, the following sets play a role in our discussion:
\[
 \mathcal{I} \equiv \big\{ p \in \mathcal{M} \; \rvert   \;\;
 |\tau(p)|<1, \; \rho(p)=0\big\}, 
\]
corresponding to the \emph{cylinder at spatial infinity},
and 
\[
 \mathcal{I}^{+} \equiv \big\{ p\in \mathcal{M} \; \rvert \; \tau(p)=1, \; \rho(p)=0
  \big\}, \qquad \mathcal{I}^{-} \equiv \big\{p \in \mathcal{M}\; \rvert \; \tau(p)=-1, \; \rho(p)=0\big\},
\]
which describe the \emph{critical sets} where null infinity touches
spatial infinity.
Additionally, let
\begin{equation*}\tilde{\mathcal{S}}_\star=\{ p \in \mathbb{R}^4 \; \rvert \; \tilde{t}(p)=0\}, \qquad {\mathcal{S}}_\star=\{ p \in \mathcal{M} \; \rvert \; \tau(p)=0\}, \end{equation*}
describing the time symmetric hypersurface of the Minkowski spacetime. The region where ${\mathcal{S}}_\star$ intersect $\mathcal{I}$ is denoted with $\mathcal{I}^0$.

\begin{figure}[t]
\centering
\includegraphics[width=0.7\textwidth]{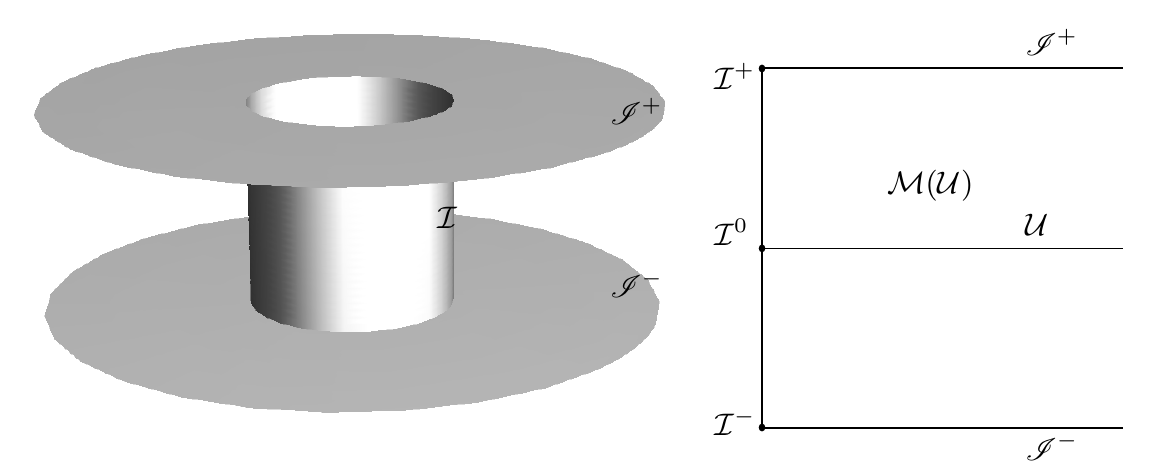}
\caption{Left: schematic representation of the cylinder at spatial
infinity of the Minkowski spacetime in the so-called \emph{F-gauge} where null
infinity corresponds to the locus of points with $\tau=\pm 1$. The
cylinder $\mathcal{I}$ is a total characteristic of Maxwell-scalar
field equations. Right: longitudinal section in which the angular
dependence has been suppressed. Here $\mathcal{U}$ denotes an open set
in a neighbourhood of $i$ and $\mathcal{M}(\mathcal{U})$ its
development; $\mathcal{I}^\pm$ are the critical sets where the
cylinder meets spatial infinity and $\mathcal{I}^0$ is the
intersection of the cylinder with the initial hypersurface. These
figures are coordinate rather than conformal diagrams ---in
particular, conformal geodesics correspond to vertical lines. Figure
taken from Figure 20.3 in page 543 of \cite{CFEBook}. \textcopyright
 Cambridge University Press. Reprinted with permission.}
\label{Figure:CylinderHorizontal}
\end{figure}

\section{The Maxwell-scalar field system}
\label{Section:MSFEqns}
In this section we provide a brief account of the Maxwell-scalar field
system with particular attention to its conformal properties and
formulation in terms of spinors. 

\subsection{Equations in the physical spacetime}
In the following let $\tilde{F}_{ab}$ denote an antisymmetric tensor
(\emph{the Faraday tensor}) over a spacetime
$(\tilde{\mathcal{M}},\tilde{\bmg})$ and let $\tilde{\nabla}$ be the
Levi-Civita connection of the metric $\tilde\bmg$. The Maxwell
equations with source are given by
\begin{subequations}
\begin{eqnarray}
&& \tilde{\nabla}^a \tilde{F}_{ab} = \tilde{J}_b, \label{eq:InHomMaxEq} \\ 
&& \tilde\nabla_{[a}\tilde{F}_{bc]} = 0. \label{eq:HomMaxEq}
\end{eqnarray}
\end{subequations}
The homogeneous equation \eqref{eq:HomMaxEq} is automatically
satisfied if one sets
\[
\tilde{F}_{ab} = \tilde{\nabla}_a \tilde{A}_b- \tilde{\nabla}_b \tilde{A}_a,
\]
where $\tilde{A}_a$ denotes the \emph{4-vector gauge potential}. 
Coupled to the above, we consider the \emph{conformally invariant wave
  equation}
\begin{equation}
\tilde{\mathfrak{D}}_a \tilde{\mathfrak{D}}^a \tilde{\phi}
-\frac{1}{6}\tilde{R} \tilde{\phi}=0,
\label{PhysicalWaveEqn}
\end{equation}
where $\tilde{\phi}$ denotes a complex scalar field. The coupling
between the Maxwell field and the scalar field is encoded in the covariant derivative
\[
\tilde{\mathfrak{D}}_a = \tilde{\nabla}_a - \mbox{i} \mathfrak{q} \tilde{A}_a,
\]
where $\mathfrak{q}$ is a coupling constant (the \emph{charge}). The
current $\tilde{J}_a$ in the inhomogeneous equation
\eqref{eq:InHomMaxEq} is given by
\[
\tilde{J}_a = \mbox{i}\mathfrak{q}\bigg( \overline{\tilde{\phi}} \tilde{\mathfrak{D}}_a \tilde{\phi} - \tilde{\phi} \overline{(\tilde{\mathfrak{D}}_a \tilde{\phi})}\bigg)
\]

\subsubsection{Gauge invariance}
The Maxwell-scalar field system
\eqref{eq:InHomMaxEq}-\eqref{eq:HomMaxEq} and \eqref{PhysicalWaveEqn} is invariant under the \emph{gauge
  transformation}
\begin{equation}
\tilde\phi \rightarrow \tilde\phi' = e^{\mathrm{i}\chi} \tilde\phi, \qquad
\tilde{A}_a\rightarrow \tilde{A}'_a = \tilde{A}_a +\nabla_a \chi, 
\label{GaugeTransformation}
\end{equation}
in the sense that $\tilde{F}_{ab}$ and $\tilde{J}_a$ are not affected by the
transformation. Moroever, the \emph{Lorenz gauge condition} 
\begin{equation}
\tilde{\nabla}^a
\tilde{A}_a =0,
\label{LorenzGauge}
\end{equation}
 is preserved by the transformation \eqref{GaugeTransformation}
for any $\chi$ such that 
\[
\square \chi =0.
\]
 Even with the Lorenz gauge condition imposed, there is some
residual gauge freedom left. This residual gauge freedom can be fixed
at the level of the initial conditions ---in particular, there is a
natural choice which allows to control the initial value of the  components of $A_a$ and
its derivatives by the energy of the system ---see Section
\ref{Section:InitialConditions}.
 
\subsubsection{Conformal transformation properties}

Consider a conformal rescaling of the form
\[
g_{ab}= \Xi^2 \tilde{g}_{ab}.
\]
Associated to the latter we define the \emph{unphysical Faraday tensor},
\emph{unphysical vector potential} and the 
\emph{unphysical scalar field} via
\[
F_{ab} \equiv \tilde{F}_{ab}, \qquad A_a\equiv \tilde{A}_a, \qquad \phi \equiv \Xi^{-1}\tilde{\phi},
\]
so that a computation using the standard conformal transformation
formulae (see e.g. \cite{CFEBook}) shows that
\begin{subequations}
\begin{eqnarray}
&& \nabla^a F_{ab} = J_b, \label{UnphysicalMaxwell1}\\ 
&& \nabla_{[a} F_{bc]} = 0, \label{UnphysicalMaxwell2}\\
&& F_{ab} = \nabla_a A_b- \nabla_b A_a, \label{DefinitionFaraday}\\
&& \mathfrak{D}_a\mathfrak{D}^a \phi - \frac{1}{6}R \phi=0, \label{UnphysicalWaveEquation}
\end{eqnarray}
\end{subequations}
where
\[
\mathfrak{D}_a \equiv  \nabla_a - \mbox{i} \mathfrak{q} A_a
\]
and
\begin{equation}
\label{Current}
J_a = \mbox{i} \mathfrak{q}\bigg( \overline{\phi} \mathfrak{D}_a \phi - \phi \overline{(\mathfrak{D}_a \phi)}\bigg).
\end{equation}
In particular, it follows that
\[
\tilde{J}_a = \Xi^2 J_a. 
\]
Moreover, one can verify that
\[
\nabla^a J_a =0.
\]
Introducing the Hodge dual $F^*_{ab}$ of the Faraday tensor in the
usual way via
\[
F^*_{ab} \equiv \frac{1}{2}\epsilon_{ab}{}^{cd}F_{cd},
\]
the Maxwell equation \eqref{UnphysicalMaxwell2} can be rewritten as
\begin{equation}
\nabla^aF^*_{ab}=0. \label{UnphysicalMaxwell2ALT}
\end{equation}

\subsection{Spinorial expressions}
In this subsection we provide the spinorial version of the equations
in the unphysical spacetime.

\medskip
Let $F_{AA'BB'}$ denote the spinorial counterpart of the Faraday
tensor $F_{ab}$. It satisfies the well-know decomposition
\[
F_{AA'BB'} = \phi_{AB}\epsilon_{A'B'} + \bar{\phi}_{A'B'} \epsilon_{AB},
\]
where $\phi_{AB}=\phi_{(AB)}$ is the so-called \emph{Maxwell
  spinor} ---see e.g. \cite{Ste91,CFEBook}. A calculation with this expression shows that equations
\eqref{UnphysicalMaxwell1} and \eqref{UnphysicalMaxwell2ALT} are
equivalent to
\begin{equation}
\nabla^B{}_{A'} \phi_{AB} = J_{AA'}, \label{MaxwellSpinorial}
\end{equation}
where 
\[
J_{AA'} \equiv \mbox{i} \mathfrak{q} \bigg( \bar{\phi}\nabla_{AA'}\phi
-\phi\nabla_{AA'}\bar{\phi} \bigg) + 2\frak{q}^2 |\phi|^2 A_{AA'},
\]
is the spinorial counterpart of the current $J_a$ and $A_{AA'}$ is the
spinorial counterpart of the vector potential $A_a$. Observe that both
$A_{AA'}$ and 
$J_{AA'}$ are Hermitian spinors.  In view of its symmetries, equation \eqref{DefinitionFaraday} can be
rewritten as
\begin{equation}
\phi_{AB} = \nabla_{A'(A} A_{B)}{}^{A'}. \label{SymmetrisedDerivativeA}
\end{equation}

\subsubsection{The wave equation for the vector potential and the
  generalised Lorenz gauge}
It is well known that in the Lorenz gauge, the vector potential
satisfies a wave equation. In light of the Lorenz gauge condition in
spinorial form
\begin{equation}
\nabla^{AA'}A_{AA'}=0,
\label{GLG}
\end{equation}
 it is possible
to remove the symmetrisation in equation
\eqref{SymmetrisedDerivativeA} so as to obtain
\begin{equation}
\nabla_{AA'}A_B{}^{A'}=\phi_{AB}.
\label{UnsymmetrisedDerivativeA}
\end{equation}
Applying $\nabla^A{}_{B'}$, using the spinorial Maxwell equation
\eqref{MaxwellSpinorial} and making use of the commutator of the
covariant derivative $\nabla_{AA'}$ one obtains
\[
\square A_{BB'} + 2\Phi_{AA'BB'}A^{AA'}=J_{BB'}.
\]
Now, since 
\[
J_{BB'}=2\mathfrak{q}^2 |\phi|^2 A_{BB'} +
\mbox{i}\bar{\phi} \nabla_{BB'}\phi -\mbox{i} \mathfrak{q}\phi
\nabla_{BB'}\bar{\phi}
\]
the wave equation for the vector potential reads as
\[
\square A_{BB'} + 2\Phi_{AA'BB'}A^{AA'} =2\mathfrak{q}^2 |\phi|^2 A_{BB'} +
\mbox{i}\mathfrak{q} \bar{\phi} \nabla_{BB'}\phi -\mbox{i} \mathfrak{q}\phi
\nabla_{BB'}\bar{\phi}.
\]

\subsubsection{The wave equation for the Maxwell spinor}
The unphysical charged wave equation is given by 
\[ 
g^{ab}\mathfrak{D}_a\mathfrak{D}_b \phi- \frac{R \phi}{6}=0. 
\]
This equation can be recast in spinor formalism by replacing 
\[
\mathfrak{D}_a = \nabla_a - \mbox{i} \mathfrak{q} A_a
\]
and then by separating the soldering forms so that we have
\[
\square \phi = \mathfrak{q}^2 \phi A_{AA'}A^{AA'} + 2\mathrm{i} \mathfrak{q} A^{AA'}
   \nabla_{AA'}\phi + \mathrm{i} \mathfrak{q}  \phi  \nabla_{AA'}A^{AA'} .
\]
Hence, by using the Lorenz gauge condition \eqref{GLG} we have
\[
\square \phi = \mathfrak{q}^2 \phi A_{AA'}A^{AA'} + 2\mathrm{i} \mathfrak{q} A^{AA'}
   \nabla_{AA'}\phi.
\]

\subsubsection{Summary}
In summary, the study of the Maxwell-scalar field system can be
reduced, making use of the generalised Lorenz gauge condition
\eqref{GLG}, to the system of wave equations
\begin{subequations}
\begin{eqnarray}
&& \square \phi = \mathfrak{q}^2 \phi A_{AA'}A^{AA'} + 2\mathrm{i} \mathfrak{q} A^{AA'}
   \nabla_{AA'}\phi, \label{MSFWave1}\\ 
&&\square A_{AA'} + 2\Phi_{ABA'B'} A^{BB'} = 2 \mathfrak{q} |\phi|^2 A_{AA'} +
\mbox{i}\mathfrak{q}\bar{\phi} \nabla_{AA'}\phi -\mbox{i} \mathfrak{q} \phi
\nabla_{AA'}\bar{\phi}. \label{MSFWave2}
\end{eqnarray}
\end{subequations}
These equations are supplemented by initial conditions for the values
of $\phi$ and $A_{AA'}$ and of their normal derivatives. This will be
discussed in more detail in Section \ref{Section:InitialConditions}. 

\subsection{Decomposition of the equations in the space-spinor
  formalism}
\label{Section:SpaceSpinors}

Before providing a detailed decomposition of the equations
\eqref{MSFWave1}-\eqref{MSFWave2}, it is convenient to provide a
rougher decomposition which brings to the foreground the structural
properties of the evolution system and its relation to the Maxwell
constraint equations. This decomposition is done using the
\emph{space-spinor formalism} as described in e.g. \cite{CFEBook}
---see also \cite{Ash91,Som80}. 

\subsubsection{Basic relations}
Let $\tau^{AA'}$ denote the spinorial counterpart of a timelike vector
field $\tau^a$ tangent to a congruence of curves. The Hermitian spinor
$\tau^{AA'}$ is chosen to have the normalisation
\[
\tau_{AA'}\tau^{AA'}=2.
\]
Consistent with the latter, we consider a spin dyad
$\{o^A,\,\iota^A \}$ chosen so that 
\[
\tau^{AA'} = o^A \bar{o}^{A'} +\iota^A\bar{\iota}^{A'}.
\]
It follows then that
\[
\tau_{AA'}\tau^{BA'} = \delta_A{}^B.
\]
The above relations induce a Hermitian conjugation operation via the
relation
\[
\mu^\dagger_A \equiv \tau_A{}^{A'}\bar{\mu}_{A'},
\]
with the obvious extension to higher valence spinors. In particular,
one has that $\iota_A =o^\dagger_A$. The space-spinor allows to work
with spinors with unprimed indices. In this spirit one has the following
decompositions for the spinorial counterpart of the current vector and
the vector potential:
\begin{subequations}
\begin{eqnarray}
&& J_{AA'} = \tfrac{1}{2} j \tau_{AA'} - j_{AB} \tau^B{}_{A'}, \label{JDecompostion}\\
&& A_{AA'} = \tfrac{1}{2}  \alpha \tau_{AA'} - \alpha_{AB}
   \tau^B{}_{A'}, \label{ADecomposition}
\end{eqnarray}
\end{subequations}
with $j_{AB}$ and $\alpha_{AB}$ symmetric spinors. 

\medskip
\noindent
\textbf{Decomposition of the covariant derivative.} The spinor
$\tau^{AA'}$ also induces a decomposition of the spinorial covariant
derivatives $\nabla_{AA'}$. For this, one defines the differential
operators
\[ \mathcal{D} \equiv \tau^{AA'}\nabla_{AA'}, \qquad \mathcal{D}_{AB}
\equiv \tau_{(A}{}^{A'} \nabla_{B)A'}.
\] The covariant derivatives $\mathcal{D}$ and $\mathcal{D}_{AB}$
correspond, respectively, to the \emph{Fermi} and \emph{Sen
connections} associated to the congruence defined by $\tau^a$.

\medskip
\noindent
\textbf{The Maxwell equations in space-spinor form.} Some manipulations show
that the spinorial Maxwell equation \eqref{MaxwellSpinorial} can be decomposed as
\begin{eqnarray*}
&& \mathcal{D}^{AB}\phi_{AB}= \tfrac{1}{2}j, \\
&& \mathcal{D}\phi_{AB} - 2 \mathcal{D}^Q{}_{(A}\phi_{B)Q} = -j_{AB}. 
\end{eqnarray*}
The former equation is to be interpreted as a \emph{constraint} while
the latter as \emph{evolution equations} ---in fact, it can be shown
to be (up to some numerical factors) a symmetric hyperbolic system for
the independent components of $\phi_{AB}$, see
\cite{CFEBook}. Similarly, from equation
\eqref{UnsymmetrisedDerivativeA} one obtains the system
\begin{eqnarray*}
&& \mathcal{D} \alpha + 2 \mathcal{D}^{AC}\alpha_{AC} + \sqrt{2}\alpha
   \chi^{AC}{}_{AC} -2 \sqrt{2} \alpha^{AC} \chi_A{}^B{}_{CB} -
   \chi^{AC}\alpha_{AC} =2 \mathcal{A}(x), \\
&& \mathcal{D} \alpha_{CD} -\mathcal{D}_{CD}\alpha -\sqrt{2}\alpha
   \chi_{(C}{}^A{}_{D)A} -2 \mathcal{D}_{(C}{}^A\alpha_{D)A}-
   \chi_{(C}{}^A \alpha_{D)A} \\
&& \hspace{5cm}+ 2\sqrt{2} \alpha^{AB}\chi_{(C|A|D)B}
   +\tfrac{1}{2}\alpha \chi_{CD}=2 \hat{\phi}_{CD},
\end{eqnarray*}
where $\chi_{AB}$ and $\chi_{ABCD}$ are, respectively, the \emph{acceleration}
and \emph{Weingarten spinors} defined by the relation
\[
\nabla_{AA'}\tau_{CC'} =-\tfrac{1}{2}\chi_{CD}\tau_{AA'} \tau^D{}_{C'}
+\sqrt{2}\chi_{ABCD} \tau^B{}_{A'}\tau^D{}_{C'}.
\]
and where 
\[
\hat{\phi}_{AB} \equiv \tau_A{}^{C'}\tau_B{}^{D'} \bar{\phi}_{C'D'}
\]
is the \emph{Hermitian conjugate} of $\phi_{AB}$. 

\medskip
\noindent
\textbf{The scalar field.} It is also illustrative to
express equation \eqref{MSFWave1} in terms of the Fermi and the Sen
connections $\mathcal{D}$  and $\mathcal{D}_{AB}$. Making use of the decomposition
\[
\nabla_{AA'} = \tfrac{1}{2}\tau_{AA'}\mathcal{D} - \tau^Q{}_{A'} \mathcal{D}_{AQ},
\]
a calculation gives that
\[
\mathcal{D}^2\phi + 2 \mathcal{D}_{AB}\mathcal{D}^{AB} =
-\sqrt{2}\chi^{AB}{}_{AB} \mathcal{D}\phi +
\chi^{AB}\mathcal{D}_{AB}\phi + 2\sqrt{2}\chi_A{}^Q{}_{BC} \mathcal{D}^{AB}\phi.
\]

\begin{remark}
{\em The equations presented in this section are completely general and
  make no assumption on the background spacetime. When evaluated on a
  the conformal representation of the Minkowski spacetime discussed in
Section \ref{TheCylinderAtSpatialInfinity} they acquire a much simpler form.}
\end{remark}

\subsubsection{Detailed decomposition in conformal Minkowski}
\label{Subsubsection:DetailedMinkowski}

In this section we particularise the decomposition of the various
fields to the case of the conformal representation of Minkowski
spacetime discussed in Section \ref{TheCylinderAtSpatialInfinity}.  

\medskip
\noindent
\textbf{Frame choice.} Following \cite{Fri03b} we
consider a so-called Newman-Penrose (NP) frame satisfying
\[
\bmg(\bme_{\bmA\bmA'},\bme_{\bmB\bmB'})=\epsilon_{\bmA\bmB}\epsilon_{\bmA'\bmB'},
\]
of the form
\begin{eqnarray*}
&& \bme_{\bmzero\bmzero'} =\frac{1}{\sqrt{2}}\bigg( (1-\tau)
   \bmpartial_\tau +\rho\bmpartial_\rho \bigg), \\
&& \bme_{\bmone\bmone'} =\frac{1}{\sqrt{2}}\bigg( (1+\tau)
   \bmpartial_\tau -\rho\bmpartial_\rho \bigg), \\
&& \bme_{\bmzero\bmone'} = -\frac{1}{\sqrt{2}}\bmX_+, \\
&& \bme_{\bmone\bmzero'} = -\frac{1}{\sqrt{2}}\bmX_-,
\end{eqnarray*}
where $\bmX_+$ and $\bmX_-$ are vectors spanning the tangent space of
$\mathbb{S}^2$ with dual covectors $\bmalpha_+$ and $\bmalpha_-$ such
that metric of the 2-sphere is given by
\[
\bmsigma = 2(\bmalpha^+\otimes\bmalpha^-+ \bmalpha^-\otimes\bmalpha^+).
\]

\medskip
The vector $\tau^a$ giving rise to the space-spinor decomposition of
the Maxwell-scalar field introduced in Section
\ref{Section:SpaceSpinors} is chosen as
\[
\tau^a = e_{\bmzero\bmzero'}{}^a+e_{\bmone\bmone'}{}^a =\sqrt{2} (\bmpartial_\tau)^a.
\]

\medskip
A peculiarity of the conformal representation introduced in equation
\eqref{MetricCylinder} is that the Ricci scalar vanishes ---that is,
\[
R[\bmg]=0.
\]

\medskip
\noindent
\textbf{The reduced wave equations.}
As the expression of the wave operator in the F-gauge acting on a spin-weighted scalar will
be used repeatedly , it is convenient to define the
\emph{F-reduced wave operator} $\blacksquare$ acting on a scalar
$\zeta$ as
\begin{equation}
\blacksquare \zeta \equiv (1-\tau^2) \ddot\zeta +
2\tau\rho \dot{\zeta}' -\rho^2\zeta''
-2\tau \dot\zeta -\frac{1}{2}\eth \bar{\eth}\zeta
-\frac{1}{2}\bar{\eth}\eth \zeta,
\label{FReducedWaveOperator}
\end{equation}
where for simplicity of the presentation we have used the notation
\[
\dot{\phantom{X}}\equiv \partial_\tau, \qquad \phantom{X}^\prime\equiv \partial_\rho,
\]
and $\eth$ and $\bar\eth$ denote the NP \emph{eth} and \emph{eth bar}
operators ---see e.g. \cite{PenRin84,Ste91}. In particular, the
operator $\tfrac{1}{2}(\eth \bar{\eth}
-\bar{\eth}\eth)$ corresponds to the Laplacian on $\mathbb{S}^2$.

\medskip
After some lengthy computations, best carried out using the suite
{\tt xAct} for tensorial and spinorial manipulations in the {\tt
  Wolfram programming language} \cite{xAct}, the wave equations
\eqref{MSFWave1}-\eqref{MSFWave2} can be seen to be equivalent to the
scalar system:
\begin{eqnarray*}
&& \blacksquare \phi = \mathfrak{q}^2\phi \bigg( \frac{1}{2}\alpha^2 -2\alpha_1^2
   +2 \alpha_0\alpha_2\bigg) + \mathrm{i}\sqrt{2}\mathfrak{q} \big( 2\alpha_1
   \rho\phi' + \alpha\dot{\phi} -2\tau \alpha_1 \dot{\phi}
   +\alpha_0\eth\phi-\alpha_2\bar\eth\phi\big), \\ 
&& \blacksquare \alpha -4 \dot{\alpha}_1 = 2\mathfrak{q}^2 \alpha |\phi|^2 +
\mbox{i}\sqrt{2} \mathfrak{q} \big( \dot{\phi} \bar{\phi} - \dot{\bar{\phi}}\phi
   \big), \\
&& \blacksquare\alpha_0
   +\alpha_0=2\mathfrak{q}^2 \alpha_0 |\phi|^2 +
   \frac{\mathrm{i}\mathfrak{q}}{\sqrt{2}}(\phi\bar\eth\bar\phi -\bar\phi\bar\eth\phi),\\
&&  \blacksquare\alpha_1 -\dot{\alpha}=2\mathfrak{q}^2 \alpha_1 |\phi|^2 +
   \frac{\mathrm{i}\mathfrak{q}\rho}{\sqrt{2}}(\phi\bar{\phi}'-\bar{\phi}\phi')+
   \frac{\mathrm{i}\mathfrak{q}\tau}{\sqrt{2}}(\bar{\phi}\dot{\phi}-\phi\dot{\bar{\phi}}),
  \\
&& \blacksquare\alpha_2
   +\alpha_2=2\mathfrak{q}^2 \alpha_2 |\phi|^2 +
   \frac{\mathrm{i}\mathfrak{q}}{\sqrt{2}}(\bar\phi\eth\phi -\phi\eth\bar\phi).
\end{eqnarray*}
The scalars $\alpha$, $\alpha_0$, $\alpha_1$ and $\alpha_2$ have spin
weight $0, \; 1,\; 0,\; -1$, respectively. In particular, $\alpha$ denotes the \emph{time component} of
the Hermitian spinor $A_{AA'}$ while $\alpha_0,\; \alpha_1,\;
\alpha_2$ are the independent components of its \emph{spatial part}
---cfr. the decomposition in equation \eqref{ADecomposition}. 

\medskip
It is observed that the righthand sides
of the above wave equations can be decoupled if one defines
\[
\alpha_\pm \equiv \alpha \pm \alpha_1, \qquad j_\pm \equiv j \pm  j_1,
\]
all of them of spin-weight $0$. In terms of these new variables the system of wave equations can be
rewritten as
\begin{eqnarray*}
&& \blacksquare \phi = s, \\ 
&& \blacksquare \alpha_+ -2 \dot{\alpha}_+ = j_+, \\ 
&& \blacksquare \alpha_- +2\dot{\alpha}_- = j_-, \\ 
&& \blacksquare \alpha_0 +\alpha_0 = j_0, \\ 
&& \blacksquare \alpha_2 +\alpha_2 = j_2, 
\end{eqnarray*}
with the obvious definitions.  For future use, it is observed that the source terms
\begin{eqnarray*}
&s=s(\bar{x},\phi,\alpha_0,\alpha_2,\alpha_\pm), &\\
&j_0=j_0(\bar{x},\phi,\alpha_0,\alpha_2,\alpha_\pm), \qquad
j_2=j_2(\bar{x},\phi,\alpha_0,\alpha_2,\alpha_\pm), \qquad j_\pm=j_\pm(\bar{x},\phi,\alpha_0,\alpha_2,\alpha_\pm),&
\end{eqnarray*}
are at most cubic polynomial expression in the unknowns $\phi,\, \alpha_0,\,
\alpha_2,\,\alpha_\pm$. 

\section{General structural
  properties and expansions near spatial infinity}
\label{StructuralProperties}

In this section we discuss general structural properties of the
evolution system, equations \eqref{NonlinearEvolutionSystemSchematic1}-\eqref{NonlinearEvolutionSystemSchematic5}, associated to the
Maxwell-scalar field system. In particular, we study a type of
asymptotic expansions near spatial infinity which was first
introduced, for the conformal Einstein field equations, by H. Friedrich in
\cite{Fri98a}. In the following, we refer to these expansions as
\emph{F-expansions}. This construction exploits the fact that the
cylinder at spatial infinity, $\mathcal{I}$,  introduced in Section \ref{TheCylinderAtSpatialInfinity} is a
\emph{total characteristic} of the evolution equations associated to
the Maxwell-scalar system.  Accordingly, the evolution equations
reduce to an interior system (\emph{transport equations}) upon
evaluation on the cylinder $\mathcal{I}$. This can clearly seen from the form of the reduced wave
operator $\blacksquare$ as given by equation
\eqref{FReducedWaveOperator} ---all the $\partial_\rho$ derivatives
disappear from the equation if one sets $\rho=0$. These
transport equations allow to relate properties of the initial data,
as defined on a fiduciary initial hypersurface $\mathcal{S}_\star$,
with radiative properties of the solution which are defined at null
infinity $\mathscr{I}^\pm$. 

\medskip
For convenience of the subsequent discussion we define
\[
\bmalpha \equiv
\left(
\begin{array}{c}
\alpha_+ \\
\alpha_- \\
\alpha_0 \\
\alpha_2
\end{array}
\right),
\qquad 
\bmj \equiv 
\left(
\begin{array}{c}
j_+ \\
j_- \\
j_0 \\
j_2
\end{array}
\right),
\qquad
\mathbf{A}\equiv 
\left(
\begin{array}{cccc}
-2 & 0 & 0 & 0\\
0 & 2 & 0 &0 \\
0 & 0 & 0 & 0 \\
0 & 0 & 0 & 0
\end{array}
\right),
\qquad
\mathbf{B}\equiv 
\left(
\begin{array}{cccc}
0 & 0 & 0 & 0\\
0 & 0 & 0 &0 \\
0 & 0 & 1 & 0 \\
0 & 0 & 0 & 1
\end{array}
\right).
\] 
In terms of the above vectors an matrices the system
\eqref{NonlinearEvolutionSystemSchematic1}-\eqref{NonlinearEvolutionSystemSchematic5}
can be rewritten as
\begin{subequations}
\begin{eqnarray}
&& \blacksquare \phi = s, \label{MatricialWave1}\\
&& \blacksquare \bmalpha + \mathbf{A} \dot\bmalpha +
   \mathbf{B}\bmalpha = \bmj. \label{MatricialWave2}
\end{eqnarray}
\end{subequations}
The source terms can, in turn, be written as
\begin{eqnarray*}
&& s = \phi \bmalpha^\dagger \mathbf{Q} \bmalpha + \bmalpha^\dagger
   \mathbf{K}(x) \bmpartial\phi, \\
&& \bmj = 2\mathfrak{q} |\phi|^2 \bmalpha + \phi \mathbf{L} \bmpartial\bar{\phi}+ \bar{\phi}\mathbf{L}\bmpartial\phi,
\end{eqnarray*}
where ${}^\dagger$ denotes the Hermitian transpose of a vector
(i.e. transposition plus complex conjugation), $\mathbf{Q}$ is a
constant matrix, while $\mathbf{K}(x)$ and $\mathbf{L}(x)$ are coordinate-dependent
matrices. Observe that if $\mathfrak{q}=0$ then  $\mathbf{K}(x)=\mathbf{L}(x)=0$. 

\subsection{Transport equations on the cylinder at spatial infinity}
The key observation in our analysis is that the F-reduced wave
operator $\blacksquare$, as defined by \eqref{FReducedWaveOperator},
reduces to an operator intrinsic to $\mathcal{I}$. Defining the
\emph{F-reduced wave operator on $\mathcal{I}$}, $\blacktriangle\equiv
\blacksquare|_{\mathcal{I}}$, acting on a scalar
$\zeta$ as
\[
\blacktriangle \zeta \equiv (1-\tau^2) \ddot\zeta
-2\tau \dot\zeta -\frac{1}{2}(\eth \bar\eth + \bar\eth \eth)\zeta.
\]
one readily observes that this operator is intrinsic to
$\mathcal{I}$. An alternative way of expressing this observation is
that the cylinder at spatial infinity is a total characteristic of the
evolution system
\eqref{MatricialWave1}-\eqref{MatricialWave2}. 

\begin{remark}
{\em The intrinsic operator $\blacktriangle$ is clearly hyperbolic for
$|\tau|<1$. However, at $\tau=\pm1$ it degenerates. To investigate the
effect of this degeneracy at the critical sets $\mathcal{I}^\pm$ it is
convenient to study the transport equations implied by evolution equations.}
\end{remark}

\subsubsection{The leading order transport equations}
Evaluating equations
\eqref{MatricialWave1}-\eqref{MatricialWave2} on $\mathcal{I}$
(i.e. at $\rho=0$) one obtains the interior system of wave equations:
\begin{subequations}
\begin{eqnarray}
&& \blacktriangle \phi^{(0)}  = \phi^{(0)} \bmalpha^{(0)\dagger} \mathbf{Q}\bmalpha^{(0)} + \bmalpha^{(0)\dagger}
   \mathbf{K} \bmpartial\phi^{(0)}, \label{ZerothOrderTransportSchematic1}\\
&&  \blacktriangle \bmalpha^{(0)} + \mathbf{A} \dot\bmalpha^{(0)} +
   \mathbf{B}\bmalpha^{(0)} =2\mathfrak{q} |\phi^{(0)}|^2 \bmalpha^{(0)} + \phi^{(0)} \mathbf{L}
   \bmpartial\bar{\phi}^{(0)}+ \bar{\phi}^{(0)}\mathbf{L}\bmpartial\phi^{(0)}, \label{ZerothOrderTransportSchematic2}
\end{eqnarray}
\end{subequations}
with 
\[
\phi^{(0)} \equiv \phi|_{\mathcal{I}}, \qquad \bmalpha^{(0)}\equiv \bmalpha|_{\mathcal{I}}.
\]
Initial data for the system
\eqref{ZerothOrderTransportSchematic1}-\eqref{ZerothOrderTransportSchematic2}
is provided by the restriction of the initial data for the fields
$\phi$, $\dot\phi$, $\bmalpha$ and $\dot\bmalpha$ on
$\mathcal{S}_\star$ to $\mathcal{I}$. 

\begin{remark}
{\em The interior system
  \eqref{ZerothOrderTransportSchematic1}-\eqref{ZerothOrderTransportSchematic2}
  is, in principle, coupled and non-linear. However, certain classes
  of initial data allow for a decoupling of the system. This feature is
  discussed in Section \ref{Section:AsymptoticExpansions}.}
\end{remark}

\subsubsection{Higher order transport equations}
\label{Subsubsection:HigherOrderTransportEqns}
Making use of the structural properties of the evolution system
\eqref{MatricialWave1}-\eqref{MatricialWave2} it is possible to
consider higher order generalisations of the transport equations
introduced  the
previous subsection. To this end, one consider the commutator
\[
\partial_\rho^p \blacksquare \zeta - \blacksquare\partial_\rho^p \zeta =
2\tau p \partial_\rho^p\dot{\zeta} -2 p\rho \partial_\rho^p \zeta' - p(p-1)\partial_\rho^p\zeta,
\]
with $\partial^p_\rho$ denoting $p$ applications of the derivative
$\partial_\rho$. Now, applying the operator $\partial_\rho$ to
equations \eqref{MatricialWave1}-\eqref{MatricialWave2} a total number
of $p$ times and restricting to $\mathcal{I}$ one finds that 
\begin{eqnarray*}
&& \blacktriangle \phi^{(p)} + 2\tau p \dot{\phi}^{(p)} - p(p-1) \phi^{(p)} = s^{(p)}, \\
&& \blacktriangle \bmalpha^{(p)} + \mathbf{A} \dot\bmalpha^{(p)}+ 2\tau p \dot{\bmalpha}^{(p)}  +
   \mathbf{B}\bmalpha^{(p)} - p(p-1) \bmalpha^{(p)} = \mathbf{R}^{(p)},
\end{eqnarray*}
where $s^{(p)}$ and $\mathbf{R}^{(p)}$ denote source terms which
depend on the solutions of the lower order transport equations
---i.e. $(\phi^{(p')},\,\bmalpha^{(p')})$ for $p'$ such that $0\leq
p'\leq p-1$. \emph{This observation allows to implement a recursive scheme
to compute the solutions to the transport equations to any arbitrary
order ---module computational complexities.}

\section{Initial conditions}
\label{Section:InitialConditions}
In this section we discuss the construction of initial data for the
Maxwell-scalar field system in the Lorenz gauge. Accordingly,
throughout it is assumed that 
\[
\nabla^a A_a =0.
\]

\subsection{General remarks}
The wave equations \eqref{MSFWave1}-\eqref{MSFWave2} suggest that a
natural prescription of initial data for the Maxwell-scalar field
system is
\[
\phi_\star, \qquad \mathcal{D}\phi_\star, \qquad A_{AA'\star}, \qquad \mathcal{D}A_{AA'\star}.
\]
Notice that the components of $A_{AA'}$ and $\mathcal{D}A_{AA'\star}$
cannot be prescribed freely. Moreover, there is some gauge freedom that
can be used to set certain components to zero ---see below. 

\begin{remark}
{\em The above is not necessarily the most physical way of prescribing
initial conditions. A more physical choice is to prescribe 
\[
\phi_\star, \qquad \mathcal{D}\phi_\star, \qquad \eta_{AB\star},
\qquad \mu_{AB\star}, 
\]
where $\eta_{AB}$ and $\mu_{AB}$ denote, respectively, the spinorial
counterparts of the electric and magnetic parts of the Faraday tensor
respect to the normal to the initial hypersurface
$\mathcal{S}_\star$. In order to fix the asymptotic behaviour one
requires finiteness of the energy
\begin{equation}
\mathcal{E}_\star \equiv \frac{1}{2}\int_{\mathbb{R}^3} \big(
|\mathfrak{D}\phi_\star|^2 + |\bmeta_{\star}|^2+|\bmmu_{\star}|^2
\big)\mathrm{d}\mu.
\label{Definition:Energy}
\end{equation}
In addition, the electric and magnetic parts are subject to the Gauss
constraints implied by the equation
\[
\mathcal{D}^{AB}\phi_{AB}= \tfrac{1}{2}j.
\] }
\end{remark}

\subsection{Data on time symmetric hypersurfaces}
In the following we assume, for simplicity, that the initial
hypersurface $\mathcal{S}_\star$ is the time symmetric hypersurface
with $t=\mbox{constant}$ in the Minkowski spacetime. Accordingly the
extrinsic curvature vanishes in this hypersurface ---thus, we have
that in the initial hypersurface $\mathcal{D}_{AB}=D_{AB}$ ---that is,
the Sen connection coincides with the Levi-Civita connection of the
intrinsic metric to $\mathcal{S}_\star$. Notice, however,
that it is not assumed that the acceleration vanishes on the initial
hypersurface ---this is for consistency with the conformal Gaussian
gauge used to write the evolution equations. 

\medskip
Following the discussion in \cite{SelTes10} we make use of the
residual gauge freedom in the Lorenz gauge to set the initial value of
the time components of $A_a$ and $\mathcal{D}A_a$ to zero
initially. In terms of the space-spinor split of $A_a$ this is
equivalent to requiring
\[
\alpha_\star =0, \qquad \mathcal{D}\alpha_\star =0. 
\]
It follows then from the Lorenz gauge condition that 
\begin{equation}
D^{AB} \alpha_{AB} = \tfrac{1}{2}\chi^{AB}\alpha_{AB}.
\label{Constraint:DivergenceAlpha}
\end{equation}
This equation has to be treated as a constraint on the spatial
part $\alpha_{AB}$. Observe how this last equation involves the acceleration. 

\medskip
The definition of the Maxwell spinor in terms of $A_{AA'}$ yields the
condition 
\[
\phi_{AB} = -\tfrac{1}{2} \dot{\alpha}_{AB}- D_{(A}{}^Q\alpha_{B)Q},
\qquad \mbox{on} \quad \mathcal{S}_\star,
\]
where for brevity we have written $\dot{\alpha}_{AB}\equiv
\mathcal{D}\alpha_{AB}$. Substituting this relation into the Gauss
constraint
\[
D^{AB}\phi_{AB} =\tfrac{1}{2}j
\]
one concludes that 
\begin{equation}
D^{AB}\dot{\alpha}_{AB} = -j.
\label{Constraint:DivergenceDotAlpha} 
\end{equation}

\subsubsection{Solving the constraints for $\alpha_{AB}$ and
  $\dot{\alpha}_{AB}$}
In order to solve the constraint equations
\eqref{Constraint:DivergenceAlpha} and
\eqref{Constraint:DivergenceDotAlpha} one can make use of the Ansatz
\begin{subequations}
\begin{eqnarray}
&& \alpha_{AB} = D_{AB} \sigma +
   D_{(A}{}^Q\sigma_{B)Q}, \label{Ansatz:Alpha}\\
&& \dot{\alpha}_{AB} = D_{AB} \pi +
   D_{(A}{}^Q\pi_{B)Q}, \label{Ansatz:DotAlpha}
\end{eqnarray}
\end{subequations}
with $\sigma$, $\pi$ scalars and $\sigma_{AB}$, $\pi_{AB}$ symmetric,
real valence 2 spinors ---the latter is essentially a spinorial
version of the Helmholtz decomposition. The substitution of the Ansatz
\eqref{Ansatz:Alpha}-\eqref{Ansatz:DotAlpha} in the
constraints \eqref{Constraint:DivergenceAlpha} and
\eqref{Constraint:DivergenceDotAlpha} leads to elliptic equations for
the scalars $\sigma$, $\pi$. The spinors $\sigma_{AB}$ and $\pi_{AB}$
are free data. 

\subsubsection{Time symmetric initial conditions}
Time symmetric initial data conditions for the
Maxwell-scalar system, i.e. initial conditions giving rise to
solutions which are time reflection symmetric with respect to the hypersurface
$\mathcal{S}_\star$, are set by requiring 
\[
\dot{\phi} =0, \qquad \dot{\alpha}_{AB}=0, \qquad \mbox{on} \quad \mathcal{S}_\star.
\]
It follows from the above that $j=0$. Thus, the only constraint
equation left to solve is equation \eqref{Constraint:DivergenceAlpha} which can
then be solved using the Ansatz \eqref{Ansatz:Alpha}. A further
consequence of $\dot{\alpha}_{AB}=0$ is that 
\[
\phi_{AB} =- D_{(A}{}^Q \alpha_{B)Q}, \qquad \hat{\phi}_{AB} =- D_{(A}{}^Q \alpha_{B)Q}.
\]
Now, defining the \emph{electric} and \emph{magnetic parts} of
$\phi_\star$ with respect to the Hermitian spinor $\tau^{AA'}$ by 
\[
\eta_{AB} \equiv \tfrac{1}{2}\mathrm{i}(\phi_{AB} + \hat{\phi}_{AB}),
\qquad {\mu}_{AB} \equiv \tfrac{1}{2} (\phi_{AB} -\hat{\phi}_{AB}),
\]
one readily finds that 
\[
\mu_{AB}=0, \qquad \mbox{on} \quad \mathcal{S}_\star. 
\]

\begin{remark}
{\em This is the spinorial version of the well-known result stating that
the magnetic part of time symmetric data for the Maxwell field has
vanishing magnetic part.} 
\end{remark}

\subsubsection{Asymptotic conditions}
The asymptotic behaviour of the initial data can be fixed in a natural
way from the requirement of the finiteness of the energy on the
(physical) initial hypersurface.

\medskip
\noindent
\textbf{Scalar field.} The finiteness of the energy as defined by equation \eqref{Definition:Energy} requires 
\[
\tilde{\mathfrak{D}}_a \tilde{\phi}_\star = o(\tilde{r}^{-1}), \qquad
(\tilde{\eta}_{\bmA\bmB})_\star= o(\tilde{r}^{-1}), \qquad (\tilde{\mu}_{\bmA\bmB})_\star = o(\tilde{r}^{-1}).
\]
These conditions are satisfied if
\[
\tilde{\phi}_\star =  o(\tilde{r}^{-1/2}), \qquad (\tilde{\alpha}_{\bmA\bmB})_\star = o(\tilde{r}^{-1/2}).
\]
In particular, one can consider an initial scalar field with leading
behaviour given by
\[
\tilde{\phi}_\star = \mathring{\varphi}\tilde{r}^{-1}+\cdots, \qquad
\mbox{with} \quad
\mathring{\varphi} \quad \mbox{a constant}.
\]
In order to obtain the \emph{conformal} version of the above condition
recall that $\Theta=\rho(1-\tau^2)$ and, moreover, $\tilde{r}=1/\rho$ and
$\varpi =\Theta_\star =\tilde{r}^{-1}=\rho$. Thus,
\[
\phi_\star =\varpi^{-1} \tilde{\phi}_\star
=\rho^{-1}\tilde{\phi}_\star =\mathring{\varphi}+ \cdots.
\]
For simplicity one can assume that $\phi_\star$ is analytic in a
neighbourhood of spatial infinity. 

\medskip
\noindent
\textbf{Electric field.} To analyse the asymptotic behaviour of the electric part of the
Maxwell field from the conformal point of view, it is observed that 
\[
\tau^a = \varpi^{-1} \tilde{\tau}^a,
 \]
where $\tilde{\tau}^a$ and $\tau^a$ are, respectively, the physical
and unphysical normals to the initial hypersurface
$\mathcal{S}_\star$. Accordingly, for a Coulomb-type field
$\tilde{E}_i = O(1/\tilde{r}^2)$ it follows that
\[
E_i = \varpi^{-1} \tilde{E}_i = O(\rho).
\]
In terms of the components with respect to the frame one has
\[
E_\bmi =O(1), \qquad \eta_{\bmA\bmB} =O(1).
\]

\medskip
\noindent
\textbf{Gauge potential.} For the gauge potential one has that $\tilde{A}_a=A_a$. Moreover, for
a Coulomb-type field one has $\tilde{A}_a =O(1/\tilde{r})$ so that in
terms of the spatial components with respect to the frame:
\[
A_\bmi =O(1), \qquad \alpha_{\bmA\bmB} =O(1).
\]

\subsubsection{An Ansatz for the initial data}
\label{Subsubsection:AnsatzInitialData}
In order to give a more concrete Ansatz for the construction of the
initial data for the Maxwell-scalar field system, in the following it
will be  assumed that \emph{the freely specifiable data $\phi_\star$ and
$\sigma_{\bmA\bmB}$ are analytic in a neighbourhood of spatial
infinity.} It follows then from the ellipticity of the equation for the
 $\sigma$ that this scalar will also be analytic in a neighbourhood of
 $i^0$. Consistent with the above let
\begin{subequations}
\begin{eqnarray}
&& \phi_\star = \sum_{n=0}^\infty \sum_{l=0}^n \sum_{m=-1}^l
   \frac{1}{n!}\varphi_{n;l,m} (Y_{lm})\rho^n, \label{AnsatzData1}\\
&& \sigma_i = \sum_{n=|1-i|}^\infty \sum_{l=|1-i|}^n \sum_{m=-l}^l
   \frac{1}{n!}\sigma_{i,n;l,m} ({}_{1-s}Y_{lm})\rho^n, \label{AnsatzData2}
\end{eqnarray}
\end{subequations}
where $\sigma_i$ for $i=0,\, 1,\, 2$ denote the three (complex)
independent components of $\sigma_{\bmA\bmB}$ and ${}_s Y_{lm}$ denote
the spin-weighted spherical harmonics ---see
e.g. \cite{PenRin84,Ste91}. Moreover, $\varphi_{n;l,m}$ and
$\sigma_{i,n;l,m}$ are constants. Finally, in accordance to the above we look for a scalar of the form 
\begin{equation}
\sigma = \sum_{n=0}^\infty \sum_{l=0}^n \sum_{m=-1}^l
   \frac{1}{n!}\sigma_{n;l,m} (Y_{lm}) \rho^n,
\label{AnsatzData3}
\end{equation}
with $\sigma_{n;l,m}$ constants. 

\section{Solution jets}
\label{Section:AsymptoticExpansions}

In this section we start our systematic study of the solutions to the
Maxwell-scalar field in a neighbourhood of spatial infinity. In order
to gain some insight into the nature of the solution we first analyse
the \emph{decoupled case} in which the charge constant $\mathfrak{q}$
is set to zero. In this case the Maxwell field and the scalar field
decouple from each other and the resulting evolution equations are
linear. We then contrast the behaviour of this decoupled case with
that of the case where $\mathfrak{q}\neq 0$. 

\medskip
We recall that the system to be solved can be written as 
\begin{subequations}
\begin{eqnarray}
&& \blacksquare \phi = s, \label{NonlinearEvolutionSystemSchematic1}\\
&& \blacksquare \alpha_+ -2 \dot{\alpha}_+ = j_+, \label{NonlinearEvolutionSystemSchematic2}\\
&& \blacksquare \alpha_- +2\dot{\alpha}_- = j_-, \label{NonlinearEvolutionSystemSchematic3}\\
&& \blacksquare \alpha_0 +\alpha_0 = j_0, \label{NonlinearEvolutionSystemSchematic4}\\
&& \blacksquare \alpha_2 +\alpha_2 = j_2, \label{NonlinearEvolutionSystemSchematic5}
\end{eqnarray}
\end{subequations}
where the source terms $s$, $j_\pm$, $j_0$ and $j_2$ are polynomial
expressions (at most of third order) in the unknowns. 

\begin{remark}
{\em A key property of the unknowns in the system
  \eqref{NonlinearEvolutionSystemSchematic1}-\eqref{NonlinearEvolutionSystemSchematic5}
is that they all possess a well defined spin-weight ---see
e.g. \cite{PenRin84,Ste91, CFEBook}. More precisely, one has that:
\begin{eqnarray*}
&& \phi, \; \alpha_\pm \qquad \mbox{have spin-weight $0$}, \\
&& \alpha_0 \qquad \mbox{has spin-weight $1$}, \\
&& \alpha_2 \qquad \mbox{has spin-weight $-1$.}
\end{eqnarray*}
The above spin-weights determine the type of expansions of the
coefficients in terms of spin-weighted spherical harmonics.}
\end{remark}

\medskip
In order to ease the discussion we make
the following simplifying assumptions:

\begin{assumption}
\label{Assumption:InitialData}
{\em The discussion will be restricted, in first instance, to the time
symmetric case. Accordingly, it is assumed that
\[
\dot\phi_\star =0, 
\qquad \dot\alpha_\star =0, \qquad \dot\alpha_{AB\star} =0.
\]
In addition, assume initial
conditions for which
\[
\alpha_{\pm\star}|_{\mathcal{I}} =0, \qquad \alpha_{0\star}|_{\mathcal{I}} =0, \qquad \alpha_{2\star}|_{\mathcal{I}} =0.
\]
and, in general, 
\[
\phi_\star|_{\mathcal{I}} \neq 0. 
\]
}
\end{assumption}

Observe that the class of initial data to be considered has a vector
potential which, on the initial hypersurface, vanishes to leading order
at spatial infinity. Crucially, the scalar field does not vanish at
leading order. \emph{The main conclusions of our analysis can be extended, at the expense
of lengthier computations, to a more general non-time symmetric setting.}

\medskip
In the following, for conciseness we write the conditions in
Assumption \ref{Assumption:InitialData} as:
\begin{subequations}
\begin{eqnarray}
& \alpha_{\pm\star}^{(0)}
   =\alpha_{0\star}^{(0)}=\alpha_{2\star}^{(0)}=0,& \label{ZeroInitial1}\\
&   \dot\alpha_{\pm\star}^{(0)}
   =\dot\alpha_{0\star}^{(0)}=\dot\alpha_{2\star}^{(0)}=0,& \label{ZeroInitial2}\\
& \phi_\star^{(0)}= \varphi_\star, \qquad \dot\phi_\star^{(0)}=0, \label{ZeroInitial3}
\end{eqnarray}
\end{subequations}
with $\varphi_\star\in \mathbb{C}$ a constant.


\medskip
As we have seen before, see Section \ref{StructuralProperties}, the cylinder at spatial infinity is a total characteristic of
our evolution equations. We can use this property to construct, in a
recursive manner, the
\emph{jets of order $p$ at $\mathcal{I}$}, $J^p[\phi,\bmalpha]$, $p\geq
0$ of the solutions to the evolution equations. Recall that the jet is
defined as
\[
J^p[\phi,\bmalpha] \equiv \{ \partial_\rho^p\phi|_{\rho=0}, \; \partial_\rho^p\bmalpha|_{\rho=0} \}.
\]
Knowledge of the  jet $J^p[\phi,\bmalpha]$ provides very precise
information about the regularity of the solutions to the evolution
equations in a neighbourhood of spatial infinity and its relation to
the structure and properties of the initial data.

\subsection{The decoupled case}
\label{Section:DecoupledCase}
Setting the charge parameter $\mathfrak{q}=0$, equations
\eqref{NonlinearEvolutionSystemSchematic1}-\eqref{NonlinearEvolutionSystemSchematic5}
readily reduce to the linear system of equations 
\begin{eqnarray*}
&& \blacksquare \phi =0\\
&& \blacksquare \alpha_+ =0,\\
&& \blacksquare \alpha_-  =0,\\
&& \blacksquare \alpha_0 + 2\dot{\alpha}_0=0,\\
&& \blacksquare \alpha_2 - 2\dot{\alpha}_2 =0.
\end{eqnarray*}

Defining 
\[
\phi^{(p)}\equiv \partial_\rho^p \phi|_{\rho=0}, \qquad
\alpha_\pm^{(p)} \equiv \partial^p_\rho \alpha_\pm|_{\rho=0},  \qquad
\alpha_0^{(p)} \equiv \partial^p_\rho \alpha_0|_{\rho=0}, \qquad
\alpha_2^{(p)} \equiv \partial^p_\rho \alpha_2|_{\rho=0}, \qquad p\geq 0,
\]
a calculation the shows that the order elements of
$J^p[\phi,\bmalpha]$ solutions satisfy the intrinsic equations
\begin{eqnarray*}
&& \blacktriangle \phi ^{(p)}+2p\tau\dot\phi{}^{(p)}=0, \\
&& \blacktriangle \alpha ^{(p)}_+ +2p\tau\dot{\alpha}{}^{(p)}_+ = 0,\\
&&\blacktriangle \alpha ^{(p)}_- +2p\tau\dot{\alpha}{}^{(p)}_- =0,\\
&& \blacktriangle \alpha ^{(p)}_0 +2(p\tau+1) \dot{\alpha}{}^{(p)}_0 =
  0 ,\\
&& \blacktriangle \alpha ^{(p)}_2 +2(p\tau-1) \dot{\alpha}{}^{(p)}_2=  0.
\end{eqnarray*}
Accordingly, in the following we study the following three model equations:
\begin{subequations}
\begin{eqnarray}
&& \blacktriangle \zeta +2p\tau\dot\zeta=0, \label{ModelEqn1}
  \\
&&  \blacktriangle \zeta+2(p\tau+1) \dot\zeta=0, \label{ModelEqn2}
\\
&&  \blacktriangle \zeta+2(p\tau-1) \dot\zeta=0. \label{ModelEqn3}
\end{eqnarray}
\end{subequations}

\begin{remark}
\label{Remark:AnsatzExpansions}
{\em In the subsequent analysis it is assumed that:
\begin{itemize}
\item[(a)] the scalar $\zeta$ in equation \eqref{ModelEqn1} has
  spin-weight $0$ and admits an expansion of form
\[
\zeta =  \sum_{p=0}^\infty \sum_{l=0}^p \sum_{m=-1}^l
   \frac{1}{p!}\zeta_{p;l,m} (Y_{lm}) \rho^p;
\]
\item[(b)] in equation \eqref{ModelEqn2} the scalar $\zeta$ has
  spin-weight $1$ and admits an expansion of the form 
\[
\zeta=\sum_{p=1}^\infty \sum_{l=1}^p \sum_{m=-l}^l
   \frac{1}{p!}\zeta_{0,p;l,m} ({}_{1}Y_{lm})\rho^p;
\]
\item[(c)] in equation \eqref{ModelEqn3} the scalar $\zeta$ has
  spin-weight $-1$ and an expansion of the form
\[
\zeta=\sum_{p=1}^\infty \sum_{l=1}^p \sum_{m=-l}^l
   \frac{1}{p!}\zeta_{2,p;l,m} ({}_{-1}Y_{lm})\rho^p.
\]
\end{itemize}
The above expansions are consistent with the discussion regarding the
freely specifiable initial data in Section \ref{Subsubsection:AnsatzInitialData} and equations
\eqref{AnsatzData1}-\eqref{AnsatzData2} and \eqref{AnsatzData3}, in
particular. Observe that at order $\rho^p$ the highest allowed
spherical harmonic corresponds to $\ell=p$. 
}
\end{remark}

Substituting the Ans\"atze in Remark \ref{Remark:AnsatzExpansions} into
the model equations \eqref{ModelEqn1}-\eqref{ModelEqn3} one obtains,
respectively, the ordinary differential equations
\begin{subequations}
\begin{eqnarray}
&& (1-\tau^2) \ddot{\zeta}_{p;\ell,m} +2\tau (p-1) \dot{\zeta}_{p;\ell,m}
+(p+\ell)(\ell-p+1) \zeta_{p;\ell,m}=0, \label{homeq1}\\
&& (1-\tau^2)\ddot{\zeta}_{0,p;\ell,m}+2\big( (p-1)\tau +1
   \big)\dot{\zeta}_{0,p;\ell,m} +
  (p+\ell)(\ell-p+1)\zeta_{0,p;\ell,m}=0, \label{homeq2}\\
&& (1-\tau^2) \ddot{\zeta}_{2,p;\ell,m} +2\big( (p-1)\tau
   -1\big)\dot{\zeta}_{2,p;\ell,m}
   +(p+\ell)(\ell-p+1)\zeta_{2,p;\ell,m} =0.
\label{homeq3}
\end{eqnarray}
\end{subequations}
Equations \eqref{homeq1}-\eqref{homeq3} are examples of \emph{Jacobi
  ordinary differential equations}. A discussion of the theory of
these equations can be found in the monograph \cite{Sze78}. The
subsequent analysis is strongly influenced by this reference. More
details can be found in Appendix \ref{Appendix:JacobiODE}. 

\begin{remark}
{\em It can be readily verified that if $\zeta_{0,p;p,m}(\tau)$ is a
  solution to equation \eqref{homeq2} then
  $\zeta{}_{0,p;p,m}(-\tau)$ solves
  \eqref{homeq3}. Thus, it is only necessary to study two model equations.}
\end{remark}

In the decoupled case, the key insight is that the behaviour of the
solutions to equations \eqref{homeq1}-\eqref{homeq3} depends on the
value of the parameter $\ell$. For $0\leq \ell\leq p-1$, $p\geq1$ the
nature of the solutions is summarised in the following:

\begin{lemma} 
The solutions to the system \eqref{homeq1}, \eqref{homeq2} and
\eqref{homeq3} can be written as
\begin{eqnarray*}
&& \zeta_{p;\ell,m}(\tau) = A_{p;\ell,m} \left(\frac{1-\tau}{2}  \right)^p
  P_\ell^{(p,-p)}(\tau) +  B_{p;\ell,m} \left(\frac{1+\tau}{2}\right)^p P_\ell^{(-p,p)}(\tau), \\
&& \zeta_{0,p;\ell,m}(\tau) = C_{p;\ell,m} \left(\frac{1-\tau}{2}  \right)^{(p+1)}
  P_\ell^{(1+p,1-p)}(\tau) +  D_{p;\ell,m} \left(\frac{1+\tau}{2}\right)^{(p-1)} P_\ell^{(-1-p,p-1)}(\tau),\\
&& \zeta_{2,p;\ell,m}(\tau) = E_{p;\ell,m} \left(\frac{1-\tau}{2}  \right)^{(p+1)}
  P_\ell^{(1+p,1-p)}(-\tau) +  F_{p;\ell,m} \left(\frac{1+\tau}{2}\right)^{(p-1)} P_\ell^{(-1-p,p-1)}(-\tau),
\end{eqnarray*}
with $P^{(\alpha,\beta)}_n(\tau)$ Jacobi polynomials of order $n$  
and where  
\[
A_{p;\ell,m},\quad  B_{p;\ell,m}, \quad C_{p;\ell,m},\quad
D_{p;\ell,m}, \quad  E_{p;\ell,m}, \quad F_{p;\ell,m} \in \mathbb{C}
\]
 denote
some constants which can
be expressed in terms of the initial conditions.  
\end{lemma}

However, for equation \eqref{homeq1} in the case $\ell=p$ we have the following proposition:

\begin{lemma} 
\label{Lemma:ZetaLogs}
For $l=p$ the solution to the equation \eqref{homeq1} can be written as
\[
\zeta_{p;p,m}(\tau) = \left(\frac{1-\tau}{2}\right)^p \left(
  \frac{1+\tau}{2} \right)^p\left( C_{1, p;\ell,m} + C_{2, p;\ell,m} \int_0^\tau \frac{\mathrm{d}s}{(1-s^2)^{p+1}} \right), 
\]
where $C_{1, p;\ell,m},\;  C_{2, p;\ell,m}$ are integration constants.
\end{lemma}

\begin{remark}
{\em Observe that the general solution given in Lemma
  \ref{Lemma:ZetaLogs} has logarithmic singularities unless the constant $C_{2, p;\ell,m}$ vanishes.  Letting $\zeta_{\star p;p,m}\equiv \zeta_{p;p,m}(0)$ and $\dot{\zeta}_{\star p;p,m}\equiv \dot{\zeta}_{p;p,m}(0)$ one readily finds
that 
\[
C_{1, p;\ell,m}=2^{2p}\zeta_{\star p;p,m}.
\]
 Similarly, one has that 
\[
C_{2, p;\ell,m}=2^{2p}\dot{\zeta}_{\star p;p,m}.
\]
 Thus, there is no logarithmic
divergence if and only if $\dot{\zeta}_{\star p;p,m}=0$ ---that is,
\emph{when the initial data for $\zeta$ is time symmetric}. In particular
\[
\zeta_{0;0,0} = C_1 + C_2\big(\log(1-\tau) -\log(1+\tau) \big).
\]
In this case one has that the logarithmic divergences are avoided if
$\dot{\zeta}_{0;0,0}(0)=0$.}
\end{remark}

Similarly, one obtains an analogous result for equations \eqref{homeq2} and \eqref{homeq3}:

\begin{lemma}
 For $\ell=p$ the solution to equations \eqref{homeq2} and \eqref{homeq3} can be written as
\begin{eqnarray*}
&&\zeta_{0,p;p,m}(\tau) = \left(\frac{1-\tau}{2}\right)^{(p+1)} \left(
  \frac{1+\tau}{2} \right)^{(p-1)}\left( C_{3, p;\ell,m} + C_{4, p;\ell,m} \int_0^\tau \frac{\mathrm{d}s}{(1-s)^{p+2}(1+s)^p} \right), \\
&&\zeta_{2,p;p,m}(\tau) = \left(\frac{1-\tau}{2}\right)^{(p-1)} \left(
  \frac{1+\tau}{2} \right)^{(p+1)}\left( C_{5, p;\ell,m} + C_{6, p;\ell,m} \int_0^\tau \frac{\mathrm{d}s}{(1+s)^{p+2}(1-s)^p} \right),
\end{eqnarray*}
where $C_{3, p;\ell,m},\;  C_{4, p;\ell,m}, \;  C_{5, p;\ell,m}, \; C_{6, p;\ell,m}$ are integration constants.
\end{lemma}

It follows from the above that the solutions for
$\zeta_{0,p;p,m}(\tau)$ and $\zeta_{2,p;p,m}(\tau)$ have logarithmic
singularities unless the constants $C_{4, p;\ell,m}$ and $C_{6, p;\ell,m}$ vanish.
Now, if we let $\zeta_{0,\star p;p,m}\equiv \zeta_{0,p;p,m}(0)$ and
$\dot{\zeta}_{0,\star p;p,m}\equiv \dot{\zeta}_{0,p;p,m}(0)$, it follows that
\[
{\zeta}_{0,\star
  p;p,m}={\bigg{(}\frac{1}{2}\bigg{)}}^{(p+1)}{\bigg{(}\frac{1}{2}\bigg{)}}^{(p-1)}
a_\star ={\bigg{(}\frac{1}{2}\bigg{)}}^{2p}a_\star.
\]
On the other hand, we have that 
\[
\begin{split}
\dot{\zeta}_{0,p;p,m}(\tau) &=- \frac{1}{2}(p+1) \bigg{(}\frac{1-\tau}{2}\bigg{)}^{p} \bigg{(}
  \frac{1+\tau}{2} \bigg{)}^{(p-1)}\bigg{(} a_\star + \dot{a}_\star \int_0^\tau \frac{\mathrm{d}s}{(1-s)^{p+2}(1+s)^p}\bigg{)}\\
  &+ \frac{1}{2}(p-1) \bigg{(}\frac{1-\tau}{2}\bigg{)}^{(p+1)} \bigg{(}
  \frac{1+\tau}{2} \bigg{)}^{p}\bigg{(} a_\star + \dot{a}_\star \int_0^\tau \frac{\mathrm{d}s}{(1-s)^{p+2}(1+s)^p}\bigg{)}\\
  &+\bigg{(}\frac{1-\tau}{2}\bigg{)}^{(p+1)} \bigg{(}
  \frac{1+\tau}{2} \bigg{)}^{(p-1)}\frac{\dot{a}_\star }{(1-\tau)^{p+2}(1+\tau)^p}.
  \end{split}
\]
Thus, it follows that 
\[
\dot{\zeta}_{0 \star,p;p,m}= - \frac{1}{2^{(2p-1)}} a_\star +
\frac{1}{2^{2p}} \dot{a}_\star. 
\]
Hence in this case the condition $\dot{\zeta}_{0 \star,p;p,m}=0$ does
not eliminate the logarithms in the solution. However, recalling that
${\zeta}_{0\star, p;p,m}={\big{(}\frac{1}{2}\big{)}}^{2p}a_\star$, it
follows from the previous equation that
\[
\dot{a}_\star=2^{2p}\dot{\zeta}_{0 \star,p;p,m} + 2^{(2p+1)}{\zeta}_{0\star, p;p,m}.
\]
Consequently, in order to have solutions without logarithmic
divergences one needs $\dot{a}_\star=0$ or, equivalently,
\[
\dot{\zeta}_{0 \star,p;p,m} = -2{\zeta}_{0\star, p;p,m}.
\]

\begin{remark}
{\em The polynomial solutions to equation \eqref{homeq2}in the case
$\ell=p$  are, thus, of the
form
\[
\zeta_{0,p;p,m} = \zeta_{0 \star,p;p,m}(1-\tau)^{(p+1)} (
  1+\tau)^{(p-1)}.
\]
Now, since $a{}_s(\tau)\equiv a(-\tau)$ is a solution for the equation
for $\zeta_{2,p;p,m}$ we have that to avoid logarithms in the
solutions to equation \eqref{homeq3} one needs the condition 
\[
\dot{\zeta}_{2 \star,p;p,m} = 2{\zeta}_{2\star, p;p,m}.
\]
In this case, the polynomial solution is given by
\[
\zeta_{2,p;p,m}(\tau) = \zeta_{2 \star,p;p,m}(1+\tau)^{(p+1)} (
  1-\tau)^{(p-1)}.
\]}
\end{remark}

Making use of the above results, the properties of the solutions to
the transport equations implied by the decoupled (i.e. linear)
Maxwell-scalar system at the
cylinder at spatial infinity $\mathcal{I}$ can be succinctly summarised
in the following proposition:

\begin{proposition}
\label{Proposition:LogsHomEq}
Given the jet $J^p[\phi,\bmalpha]$ for $\mathfrak{q}=0$
one has that:
\begin{itemize}
\item[(i)] the elements of the jet have polynomial dependence in $\tau$ for the harmonic sectors
  with 
$0\leq \ell \leq p-1$ and, thus, they extend analytically through $\tau=\pm1$;
\item[(ii)] generically, for $\ell =p$, the solutions have logarithmic
  singularities at $\tau=\pm 1$. These logarithmic divergences can be
  precluded by fine-tuning of the initial data. 
\end{itemize} 
\end{proposition}

\begin{remark}
\label{Remark:DecoupledSummary}
{\em The key insight form the analysis of the decoupled system is that
  for a given order $p$, the elements in  $J^p[\phi,\bmalpha]$ only
  exhibit singular behaviour at the critical sets $\mathcal{I}^\pm$
  where spatial infinity touches null infinity for the harmonics with
  the highest admissible $\ell$. All other sectors with $\ell<p$ are
  completely regular for generic initial conditions.
}
\end{remark}

\subsection{The coupled case}
In this section we provide an analysis of the behaviour of the
elements of the jet $J^p[\phi,\bmalpha]$ in the case $\mathfrak{q}\neq
0$ with particular emphasis on their regularity at the critical sets
$\mathcal{I}^\pm$. In order to keep the presentation concise we focus
on the differences with the decoupled case ---see Remark \ref{Remark:DecoupledSummary}.

\subsubsection{The $p=0$ order transport equations} 
We start our analysis of the full non-linear system by looking at the
solutions corresponding to the jet $J^0[\phi,\bmalpha]$ ---that is,
the order $p=0$. Evaluating the system
\eqref{NonlinearEvolutionSystemSchematic1}-\eqref{NonlinearEvolutionSystemSchematic5}
one finds that 
\begin{subequations}
\begin{eqnarray}
&& \blacktriangle \phi ^{(0)} = s ^{(0)}, \label{ZerothOrderNonLinearSystem1}\\
&& \blacktriangle \alpha ^{(0)}_+ -2 \dot{\alpha}{}^{(0)}_+ = j ^{(0)}_+, \label{ZerothOrderNonLinearSystem2}\\
&& \blacktriangle \alpha ^{(0)}_- +2\dot{\alpha}{}^{(0)}_- = j ^{(0)}_-, \label{ZerothOrderNonLinearSystem3}\\
&& \blacktriangle \alpha ^{(0)}_0 +\alpha ^{(0)}_0 = j ^{(0)}_0, \label{ZerothOrderNonLinearSystem4}\\
&& \blacktriangle \alpha ^{(0)}_2 +\alpha ^{(0)}_2 = j ^{(0)}_2. \label{ZerothOrderNonLinearSystem5}
\end{eqnarray}
\end{subequations}
This order is non-generic as under Assumption
\ref{Assumption:InitialData}, it can be readily verified that the
above transport equations decouple and it is possible to write down
the solution explicitly. More precisely, one has that:

\begin{lemma}
\label{Lemma:ZerothOrderSolution}
The unique solution to the $0$-th order system
\eqref{ZerothOrderNonLinearSystem1}-\eqref{ZerothOrderNonLinearSystem5}
with initial conditions \eqref{ZeroInitial1}-\eqref{ZeroInitial3} is
given by 
\[
\phi^{(0)}= \varphi_\star, \qquad \alpha^{(0)}_\pm =0, \qquad
\alpha_0^{(0)}=0, \qquad \alpha^{(0)}_2=0. 
\]
\end{lemma}

\begin{remark}
{\em As it will be seen, the $0$-th order jet $J^0[\phi,\bmalpha]$
  given by the above lemma allows to start a
recursive scheme to compute the higher order jets $J^p[\phi,\bmalpha]$
with $p\geq 1$.}
\end{remark}

\subsubsection{The $p\geq1$ transport equations}
In order to analyse the properties of the jet of order $p$,
$J^p[\phi,\bmalpha]$ for given  $p=n$, we assume that we have
knowledge of the jets
\[
J^0[\phi,\bmalpha], \;\; J^1[\phi,\bmalpha], \ldots,
J^{n-1}[\phi,\bmalpha]. 
\]
Under this assumption and taking into account Lemma
\ref{Lemma:ZerothOrderSolution} one finds that the elements of
$J^p[\phi,\bmalpha]$  satisfy the equations ---cfr. the general
discussion in Subsection \ref{Subsubsection:HigherOrderTransportEqns}: 
\begin{subequations}
\begin{eqnarray}
&& \blacktriangle \phi ^{(n)}+2n\tau\dot\phi{}^{(n)}= s^{(n)}, \label{NonlinearOrderN1}\\
&& \blacktriangle \alpha ^{(n)}_+ +2(n\tau-1) \dot{\alpha}{}^{(n)}_+ =
   2\mathfrak{q}^2|\varphi_\star|^2 \alpha ^{(n)}_+ + \tilde{j}_+^{(n)}, \label{NonlinearOrderN2}\\
&&\blacktriangle \alpha ^{(n)}_-  +2(n\tau+1)\dot{\alpha}{}^{(n)}_- = 2\mathfrak{q}^2|\varphi_\star|^2 \alpha ^{(n)}_-+\tilde{j}_-^{(n)}, \label{NonlinearOrderN3}\\
&& \blacktriangle \alpha ^{(n)}_0 +2n\tau\dot{\alpha}{}^{(n)}_0+\alpha ^{(n)}_0 =
   2\mathfrak{q}^2|\varphi_\star|^2 \alpha
   ^{(n)}_0+\tilde{j}^{(n)}_0 , \label{NonlinearOrderN4}\\
&& \blacktriangle \alpha ^{(n)}_2 +2n\tau\dot{\alpha}{}^{(n)}_2+\alpha ^{(n)}_2 =  2\mathfrak{q}^2|\varphi_\star|^2 \alpha
   ^{(n)}_2 +\tilde{j}_2^{(n)}, \label{NonlinearOrderN5}
\end{eqnarray}
\end{subequations}
where $s^{(n)}$, $\tilde{j}^{(n)}_\pm$, $\tilde{j}^{(n)}_0$ and
$\tilde{j}_2^{(n)}$ depend, solely, on the elements of
$J^p[\phi,\bmalpha]$, $0\leq p\leq n-1$. 

\begin{remark}
{\em The key new feature in the above equations is the presence in
\eqref{NonlinearOrderN2}-\eqref{NonlinearOrderN5} of the terms
involving the constant $2\mathfrak{q}^2|\varphi_\star|^2$ in the
right-hand side of the equations. These terms arise from the cubic
nature of the coupling in the source terms in the Maxwell-scalar field
system. This feature does not arise in systems with quadratic coupling
like the conformal Einstein-field equations or the Maxwell-Dirac
system. In particular, observe that one is lead to consider model
homogeneous equations of the form
\begin{subequations}
\begin{eqnarray}
&& \blacktriangle \zeta +2(n\tau-1) \dot\zeta -\varkappa
   \zeta=0, \label{NonLinearModelHomogeneousEqns1}
  \\
&&  \blacktriangle \zeta+2n\tau
   \dot\zeta+(1-\varkappa)\zeta =0 \label{NonLinearModelHomogeneousEqns2}
\end{eqnarray}
\end{subequations}
with $\varkappa\equiv 2\mathfrak{q}^2|\varphi_\star|^2$. As it will be seen in
the sequel, the solutions of these equations for generic choice of
$\varkappa$ is radically different to that of the case $\varkappa=0$
---i.e. $\mathfrak{q}=0$.
}
\end{remark}

\medskip
Now, assuming that the various fields have an asymptotic expansion as
in Remark \ref{Remark:AnsatzExpansions}
one is lead to consider a hierarchy of ordinary differential equations
of the form
\begin{subequations}
\begin{eqnarray}
&&\hspace{-1.5cm} (1- \tau^2)\ddot{\phi}{}_{n; \ell,m} + 2(n-1)\tau\dot{\phi}{}_{n;
   \ell,m}+((\ell-n+1)(n+\ell)){\phi}{}_{ n; \ell,m}= s_{n;\ell,m}, \label{NonlinearODE1}\\
&&\hspace{-1.5cm}   (1- \tau^2)\ddot{\alpha}{}_{+, n; \ell,m} + 2(-1 +
   (n-1)\tau)\dot{\alpha}{}_{+, n; \ell,m}+
   (\ell(\ell+1)-n(n-1)-\varkappa){\alpha}{}_{+, n;
   \ell,m}=\tilde{j}_{+,n;\ell,m}, \label{NonlinearODE2}\\
&&\hspace{-1.5cm}  (1- \tau^2)\ddot{\alpha}{}_{-, n; \ell,m} + 2(1+ (n-1)\tau)\dot{\alpha}{}_{-, n; \ell,m}+  (\ell(\ell+1)-n(n-1)- \varkappa){\alpha}{}_{-, n; l,m}=\tilde{j}_{-,n;\ell,m}, \label{NonlinearODE3}\\
&&\hspace{-1.5cm}  (1- \tau^2)\ddot{\alpha}{}_{0, n; \ell,m} + 2(n-1)\tau \dot{\alpha}{}_{0, n; \ell,m}+((\ell-n+1)(n+\ell)-\varkappa){\alpha}{}_{0, n; \ell,m}=\tilde{j}_{0,n;\ell,m}, \label{NonlinearODE4}\\
&&\hspace{-1.5cm}   (1- \tau^2)\ddot{\alpha}{}_{2, n; \ell,m} + 2(n-1)\tau
   \dot{\alpha}{}_{2, n; \ell,m}+((\ell-n+1)(n+\ell)-
   \varkappa){\alpha}{}_{2, n; \ell,m}=\tilde{j}_{2,n;\ell,m}, \label{NonlinearODE5}
\end{eqnarray}
\end{subequations}
for $0\leq \ell \leq n$, $-\ell\leq m\leq \ell$ and 
with the \emph{source terms} 
\[
 s_{n;\ell,m}, \quad \tilde{j}_{+,n;\ell,m}, \quad
 \tilde{j}_{-,n;\ell,m}, \quad \tilde{j}_{0,n;\ell,m}, \quad \tilde{j}_{2,n;\ell,m},
\]
known as a result of the spherical harmonics decomposition of the
lower order jets $J^p[\phi,\bmalpha]$ for $0\leq p \leq n-1$. The
homogeneous version of equations
\eqref{NonlinearODE2}-\eqref{NonlinearODE5} does not fit the general
scheme of solutions discussed in Subsection \ref{Section:DecoupledCase} for the decoupled
system. In fact, one has the following general result from \cite{Sze78}
which we quote for completeness
\begin{lemma}
\label{Lemma:PolynomialSolutions}
The Jacobi ordinary differential equation
\[
(1-\tau^2) \ddot{a}
+\big(\beta-\alpha-(\alpha+\beta+2)\tau  \big) \dot{a} +\gamma a=0
\]
has polynomial solutions if and only if $\gamma$ is rational.
\end{lemma}

So, the question
is whether it is possible to characterise the solutions in an easy
manner? For
this, we resort to Frobenius's method to study the properties of the
equations in terms of asymptotic expansions at the values $\tau=\pm
1$ ---see \cite{Tes11}, Chapter 4. The homogeneous version of equation
\eqref{NonlinearODE2}-\eqref{NonlinearODE5} can be described in terms
of the model equation
\begin{equation}
 (1- \tau^2)\ddot{\zeta}+ 2(\varsigma + (n-1)\tau)\dot{\zeta}+
 (\ell(\ell+1)-n(n-1)-\varkappa){\zeta}=0
\label{CoupledHomogeneousModelODE}
\end{equation}
where 
\[
\varsigma =\left\{
\begin{array}{l}
-1 \quad  \mbox{for} \quad \alpha_+\\
1 \quad \mbox{for} \quad \alpha_- \\
0\quad \mbox{for} \quad
\phi,\;\alpha_0,\; \alpha_2
\end{array}
\right.
\]
 ---recall also that $\varkappa=2\mathfrak{q}^2
 \varphi_\star^2$. Following Frobenius's method we look for power
 series solutions of the  form
\begin{equation}
\zeta= (1- \tau)^r \sum_{k=0}^{\infty} D_k (1- \tau)^k,
\qquad D_0 \neq 0. 
\label{CoupledHomogeneousModelSolutionAnsatz}
\end{equation}
Substitution of the Ansatz
\eqref{CoupledHomogeneousModelSolutionAnsatz} into the model equation
\eqref{CoupledHomogeneousModelODE} leads to the \emph{indicial equation} 
\[
2r(r-1)-2\varsigma r-2(n-1)r=0.
\]
The solutions to the indicial equation for the various values of
$\varsigma$ are given in Table \ref{Table:SolutionIndicialEqn}.

\begin{table}[t]
\begin{center}
\begin{tabular}{c|c|c}
$\varsigma$ & $r_1$ & $r_2$\\
\hline
$-1$ & $0$ & $n-1$ \\
$0$ & $0$ & $n$ \\
$1$ & $0$ & $n+1$
\end{tabular}
\end{center}
\caption{Roots of the indicial equation.}
\label{Table:SolutionIndicialEqn}
\end{table}

Once the solutions to the indicial equation are known, Ansatz \eqref{CoupledHomogeneousModelSolutionAnsatz}
leads to a recurrence relation for the coefficients $D_k$ in the
series. The details of this computation are given in Appendix
\ref{Appendix:SeriesSolution}. The key observation for the subsequent
discussion is that for a given value of $\varsigma$, the root $r_1=0$
of the indicial polynomial does not lead to a valid series
solution as the recursion relation breaks down at some order. In
order to obtain a second, linearly independent solution to equation
\eqref{CoupledHomogeneousModelODE} one needs to consider a more
general type of Ansatz. Again, following the discussion in
\cite{Tes11} we look for a second solution of the form
\begin{equation}
\zeta=\sum_{k=0}^{\infty} G_k (1- \tau)^{k}+(1-\tau)^{r_2}\log (1- \tau) \sum_{k=0}^{\infty} M_k (1- \tau)^{k}, \qquad G_0
\neq 0,\quad M_0 \neq 0.
\label{CoupledHomogeneousModelSecondSolutionAnsatz}
\end{equation}
A detailed inspection of the recurrence relations implied by the
Ansatz \eqref{CoupledHomogeneousModelSecondSolutionAnsatz} shows that
all the coefficients $M_k$ for $k=1,\,2,\ldots$ and $G_k$ for
$k=0,\,1,\,2,\ldots$ can be expressed in terms of the coefficient
$M_0$ ---again, see Appendix
\ref{Appendix:SeriesSolution} for further details. 

\begin{remark}
{\em The previous analysis has been restricted, for concreteness, to
  the behaviour of the solutions to the homogeneous model equation
  \eqref{CoupledHomogeneousModelODE} near $\tau=1$. A similar analysis
can be carried out \emph{mutatis mutandi} to obtain the behaviour of
the solutions near $\tau=-1$. }
\end{remark}

\begin{remark}
{\em Observe that the logarithmic singularity of the solutions given
  by \eqref{CoupledHomogeneousModelSecondSolutionAnsatz} is modulated
  by a term of the form $(1-\tau)^{r_2}$. Accordingly, within the
  radius of convergence of the series, the whole solution is of class
$C^{r_2-1}$ at $\tau=\pm 1$.}
\end{remark}

\begin{remark}
{\em It is of some interest that the solutions to the model equation
  \eqref{CoupledHomogeneousModelODE} can be written in  closed form
  in terms of hypergeometric functions. This representation, however,
makes it harder to examine the regularity properties of the solutions
at the critical values $\tau=\pm 1$.}
\end{remark}

The discussion in the previous paragraphs can be summarised in the
following:

\begin{proposition}
\label{Proposition:BehaviourHomogeneousEqnsCoupledCase}
The general solution to the 
\eqref{CoupledHomogeneousModelODE} 
\[
(1- \tau^2)\ddot{\zeta}+ 2(\varsigma + (n-1)\tau)\dot{\zeta}+
 (\ell(\ell+1)-n(n-1)-\varkappa){\zeta}=0, \qquad \varkappa\neq 0
\]
with $\varsigma=-1,0,1$, $0\leq \ell \leq n$, $n=1,\,2,\,\ldots$ consists, of:  
\begin{itemize}
\item[(i)] one solution which is analytic for $\tau\in[-1,1]$;
\item[(ii)] one solution with is analytic for $\tau\in (-1,1)$ and has
  logarithmic singularities at $\tau=\pm 1$. At these singular points
  the solution is of class $C^{r_2-1}$.
\end{itemize} 
\end{proposition}

\begin{remark}
{\em The key observation from the previous analysis is the fact that
  the solutions to the homogeneous equations in the coupled case have
  one solution with logarithmic singularities for every $0 \leq \ell \leq n$
and $-\ell \leq m \leq \ell$. This is in contrast to the decoupled
case where only the solutions corresponding to the spherical harmonics
with $\ell =n$ had logarithmic divergences.} 
\end{remark}

\subsubsection{The solution to the inhomogeneous equations}
Having analysed the behaviour of the solutions to the homogeneous part
of the transport equations we proceed now to briefly discuss the
behaviour to the full inhomogeneous equations
\eqref{NonlinearODE1}-\eqref{NonlinearODE5}. For this we rely on the
method of variation of parameters as discussed in Appendix \ref{Appendix:VariationParameters}.

\medskip
In the following let $\zeta$ denote any of the unknowns
$(\phi_{n;\ell,m},\, \alpha_{+,n;\ell,m}, \, \alpha_{-,n;\ell,m},\,
\alpha_{0,n;\ell,m},\, \alpha_{2,n;\ell,m})$ in the transport
equations \eqref{NonlinearODE1}-\eqref{NonlinearODE5}. These equations
are described through the model equation 
\begin{equation}
(1- \tau^2)\ddot{\zeta} + 2(\varsigma + (n-1)\tau)\dot{\zeta}+
(n(1-n)+\ell(\ell+1)-\varkappa){\zeta}= f(\tau), \qquad
\varsigma=-1,\,0,\,1,
\label{ModelInhomogeneousEqnMain}
\end{equation}
where $f$ denotes the corresponding source terms ($s_{n;\ell,m},\,
\tilde{j}_{+,n;\ell,m},\, \tilde{j}_{-,n;\ell,m},\,
\tilde{j}_{0,n;\ell,m}, \, \tilde{j}_{2,n;\ell,m}$). 
Moreover, let $\zeta_1$ and $\zeta_2$ denote two linearly independent
solutions to the homogeneous problem. The method of variation of
parameters gives the general solution to
\eqref{ModelInhomogeneousEqnMain} in the form
\begin{equation}
  {\zeta}(\tau) =A_1(\tau){a}_1(\tau)+ A_2(\tau){a}_2(\tau),
  \label{Solution:InhomogeneousEqn}
\end{equation}
where
\begin{subequations}
\begin{eqnarray}
&& A_1(\tau) = A_{1\star} -\int_0^\tau \frac{\zeta_2(s)
   f(s)}{W_\star(1-s^2)^{n}}\bigg{(}\frac{1+s}{1-s}\bigg{)}^{2\varsigma}\mathrm{d}s,
  \label{VariationParameter1Final}\\
&& A_2(\tau) = A_{2\star} +\int_0^\tau \frac{\zeta_1(s)
   f(s)}{W_\star(1-s^2)^{n}}\bigg{(}\frac{1+s}{1-s}\bigg{)}^{2\varsigma}\mathrm{d}s, \label{VariationParameter2Final}
\end{eqnarray}
\end{subequations}
with $A_{1\star}$ and $A_{2\star}$ constants fixed by the initial
data. The details of the derivation of these expressions can be found
in Appendix \ref{Appendix:VariationParameters}. For ease of
presentation, the discussion in this subsection is focused on the the
behaviour of the solutions at $\tau=1$. A similar discussion can be
made, \emph{mutatis mutandi}, for the behaviour at $\tau=-1$. 

\medskip
Consistent with Proposition
\eqref{Proposition:BehaviourHomogeneousEqnsCoupledCase}, we distinguish two cases for the solutions of \eqref{Solution:InhomogeneousEqn} as follows:
\begin{itemize}
\item[(i)] the two solutions to the homogeneous equation are smooth at
  $\tau=1$;
  \item[(ii)] one of the solutions to the homogeneous problem is
    smooth at $\tau=1$ while the other has a logarithmic singularity.
  \end{itemize}

In the following for simplicity of the presentation it is assumed that
the source term $f$ is regular at $\tau=1$ ---i.e. it does not contain
singularities of either logarithmic type or poles.
  
  \medskip
  \noindent
  \textbf{Case (i).} We observe that the integrands in equations
  \eqref{VariationParameter1Final} and
  \eqref{VariationParameter2Final} contain a pole of order
  $n+2\varsigma$ at $\tau=1$. The decomposition in partial fractions
  will, for generic source $f(s)$, contain a term of the form
  \[
\frac{1}{1-s}
\]
which, when integrated gives rise to a logarithmic term
$\ln(1-\tau)$. This type of logarithmic singularity can be precluded
if the zeros of the expressions
\[
\zeta_1(s) f(s)(1+\tau)^{2\varsigma}, \qquad \zeta_2(s) f(s)(1+\tau)^{2\varsigma}
\]
have a very fine-tuned structure. The latter can be, in principle,
reexpressed in terms of conditions on the initial ---this task,
however, goes beyond the scope of this article. Thus, the generic
conclusion is that even if the solutions $\zeta_1$ and $\zeta_2$ to
the homogeneous problem do not contain logarithmic singularities at
$\tau=1$, the actual solutions to the transport equations at a given
order will have this type of singularities unless the initial
conditions are fine-tuned. The regularity (or more precisely, lack
thereof) of these singularities is controlled by the factors of
$(1-\tau)$ appearing in the functions $\zeta_1$ and $\zeta_2$. It can
also be readily verified that the structure of these factors in
$\zeta_1$ and $\zeta_2$ is such that the final solution as given by
formula \eqref{Solution:InhomogeneousEqn} has no poles at $\tau=1$
---that is to say, the only possible singularities are of logarithmic
type.

\medskip
  \noindent
  \textbf{Case (ii).} \emph{In the following we assume that $\zeta_2$ is the
  solution to the model homogeneous equation containing the
  logarithmic term.} A quick inspection of equation
  \eqref{VariationParameter2Final} the shows that this term will give
  rise, generically, to logarithmic singularities similar to those in
  Case (i). The situation is, however, different for expression
  \eqref{VariationParameter1Final} for which the denominator already
  contains a $\ln(1-\tau)$ term. The decomposition in terms of partial
  fractions gives rise to a term of the form
  \[
\frac{\ln(1-\tau)}{(1-\tau)},
\]
which, after being integrated, gives rise to a singular term of the form
\[
\ln^2(1-\tau).
\]
This is the most singular term arising from the integration of the
partial fractions decomposition of the integrand in
\eqref{VariationParameter1Final}. As in Case (i), the coefficients in
the partial fractions decomposition can, in principle, be expressed in
terms of initial data ---thus, this singular term could be removed by
fine-tuning. The remaining terms in the expansion give rise, at worst,
to singular terms containing $\ln(1-\tau)$ and some power of
$1-\tau$. As in Case (i), it can be verified that the solution arising
from formula \eqref{Solution:InhomogeneousEqn} does not contain poles
at $1-\tau$ ---that is, again, all singular behaviour is of
logarithmic type.

\begin{remark}
{\em More generally, in view of the recursive nature of the of the
  transport equation in which the source terms at order $n$ are given
  explicitly in terms of lower order jets, the source terms will
  contain logarithmic terms involving powers of $\ln(1-\tau)$. Because
  of the structural properties of the variation of parameters formula
  will then give rise to higher order logarithmic terms. The
  discussion in the previous paragraphs thus shows that even in the
  optimal case where the source is completely regular, logarithmic
  terms will arise.}
\end{remark}

\subsection{Summary}
The discussion in this section can be summarised in the following:

\begin{theorem}
  \label{MainTheorem:MainText}
For generic initial data for the Maxwell-scalar field the jet
$J^p[\phi,\bmalpha]$, $p\geq 1$ contains logarithmic divergences at
the $\tau=\pm 1$ ---i.e. at the critical sets $\mathcal{I}^\pm$ where
null infinity meets spatial infinity---  for \emph{all} spherical harmonic sectors. The logarithmic divergences are
of the form
\[
(1\pm \tau)^{\mu_1}\ln^{\mu_2}(1\pm\tau)
\]
for some non-negative integers $\mu_1$, $\mu_2$. 
\end{theorem}

\begin{remark}
{\em The situation described in Theorem \ref{MainTheorem:MainText} is
  to be contrasted with the situation in the decoupled case in which
  for the solution jet at order $p$,
  for generic initial data, there always exist spherical sectors
  without logarithmic singularities ---see Proposition \ref{Proposition:LogsHomEq}. Moreover, due to the absence of
  source terms the logarithmic singularities are of the form
  \[
(1\pm \tau)^{\mu_3}\ln(1\pm\tau)
\]
for some non-negative integer $\mu_3$.  It is in this sense that the
non-linear coupling of the Maxwell and scalar fields gives rise to a
more singular behaviour at the conformal boundary and, consequently, a
more complicated type of asymptotics. 
}
\end{remark}

\section{Peeling properties of the Maxwell-scalar system}
\label{Section:Peeling}
In this section we translate the results on the regularity of the
solutions of the Einstein-Maxwell at the conformal boundary obtained
in Section \ref{Section:AsymptoticExpansions} into statements about the asymptotic decay of
the fields in the physical spacetime. The most important consequence
of regularity (smoothness) at the conformal boundary of a field is the
so-called \emph{peeling} ---i.e a hierarchical decay of the various
components of, say, the Maxwell field along the generators of outgoing
light cones. As the asymptotic expansions of Section
\ref{Section:AsymptoticExpansions} generically imply a non-smooth behaviour at the
conformal boundary, one expects a \emph{modified peeling behaviour}.

\subsection{The Newman-Penrose gauge}\label{Section:NP}

The discussion of peeling properties fields is usually done in terms
of a gauge which is adapted to null infinity ---the so-called
\emph{Newman Penrose (NP) gauge}. The relation between the NP-gauge
and the F-gauge used to compute the expansions in Section
\ref{Section:AsymptoticExpansions} has been studied in detail, for the
Minkowski spacetime, in \cite{GasVal20}. In this subsection we briefly
discuss the associated transformation formulae. 

\medskip
In the following, the discussion will be restricted to the case of
$\mathscr{I}^+$.  Analogous conditions can be
formulated, \emph{mutatis mutandi}, 
for $\mathscr{I}^-$. The NP gauge is adapted to the geometry of null
infinity. Let $\{ \bme'_{\bmA \bmA'} \}$ denote a frame
satisfying $\bmeta(\bme'_{\bmA \bmA'}, \bme'_{\bmB
\bmB'})=\epsilon_{\bmA \bmB} \epsilon_{\bmA' \bmB'}$ in a
neighbourhood $\mathcal{U}$ of $\mathscr{I}^+$. The frame is said to
be in the NP-gauge if it satisfies the conditions:

\begin{itemize}
    \item[(i)] the vector $\bme'_{\bmone \bmone'}$ is tangent to
      $\mathscr{I}^+$ and is such that 
\[
    \nabla_{\bmone \bmone'} \bme'_{\bmone \bmone} =0.
\]

\item[(ii)] There exits a smooth function $u$ (\emph{retarded time}) on $\mathcal{U}$
that satisfies $\bme'_{\bmone \bmone'}(u) = 1$ at $\mathscr{I}^+$.

\item[(iii)] The vector $\bme'_{\bmzero \bmzero'}$ is required to
  satisfy  
\[
    \bme'_{\bmzero \bmzero'} = \bmeta(\mathbf{d}{u},\cdot).
\]
\item[(iv)] Let
\[
    \mathcal{N}_{u_\bullet} \equiv \{ p \in \mathcal{U} \ | \ u(p) = u_\bullet \},
\]
 where $u_\bullet$ is constant. Then the frame
$\bme'_{\bmA \bmA'}$, tangent to $\mathcal{N}_{u_\bullet} \cup
\mathscr{I}^+$, satisfies
    $$
    \nabla_{\bmzero \bmzero'} \bme'_{\bmA \bmA'} = 0 \hspace{2mm} \text{on} \hspace{2mm} \mathcal{N}_{u_\bullet}.
    $$
\end{itemize}

In \cite{GasVal20}, the relation between the NP-gauge frame $\{
\bme'_{\bmA \bmA'} \}$ and the F-gauge frame $\{ \bme_{\bmA \bmA'} \}$
for the Minkowski spacetime, as defined in Section
\ref{Subsubsection:DetailedMinkowski}, was explicitly computed. This
computation assumes the conformal factor
\[
\Theta =\rho(1-\tau^2),
\]
  and its key outcomes are summarised in the following:

\begin{proposition}
\label{Proposition:ChangeOfGauge}
The NP-gauge frame at $\mathscr{I}^+$ and F-gauge frame in the
Minkowski spacetime are related via 
\begin{equation}
    \bme'_{\bmA \bmA'} = \Lambda^{\bmB}{}_{\bmA} \bar{\Lambda}^{\bmB'}{}_{\bmA'} \bme_{\bmB \bmB'},
    \label{NP-FGauge-Relation}
\end{equation}
with
\begin{equation}
    \Lambda^{\bmzero}{}_{\bmone} = \frac{2 e^{i \omega}}{\sqrt{\rho}(1+\tau)}, \hspace{5mm} \Lambda^{\bmone}{}_{\bmzero} = \frac{e^{-i \omega} \sqrt{\rho} (1+\tau)}{2}, \hspace{5mm} \Lambda^{\bmone}{}_{\bmone} = \Lambda^{\bmzero}{}_{\bmzero} =0,
\end{equation}
where $\omega$ is an arbitrary real number that encodes the spin
rotation of the frames on $\mathbb{S}^2$. For the NP-gauge frame at
$\mathscr{I}^-$, the roles of the vectors $\bme'_{\bmzero \bmzero'}$
and $\bme'_{\bmone \bmone'}$ are interchanged, and NP-gauge frame is related to the F-gauge by equation \eqref{NP-FGauge-Relation} with $\Lambda^{\bmA}{}_{\bmB}$ given by
\begin{equation}
    \Lambda^{\bmzero}{}_{\bmone} = \frac{e^{-i \omega} \sqrt{\rho} (1-\tau)}{2} , \hspace{5mm} \Lambda^{\bmone}{}_{\bmzero} = \frac{2 e^{i \omega}}{\sqrt{\rho}(1-\tau)}, \hspace{5mm} \Lambda^{\bmone}{}_{\bmone} = \Lambda^{\bmzero}{}_{\bmzero} =0.
\end{equation}
\end{proposition}

\subsection{The scalar field}
We start our discussion of the peeling properties looking at the
scalar field. In order to carry out this computation we make the
following assumption:

\begin{assumption}
\label{Assumption:TaylorScalarField}
On $\mathcal{M}$, the scalar field $\phi$ satisfies the asymptotic
expansion
\[
\phi = \sum_{p=0}^N \frac{1}{p!}\phi^{(p)}\rho^p + o_1(\rho^N)
\]
for some sufficiently large $N$ and 
where $\phi^{(p)}$ are contained in the solution jet
$J^{(p)}[\phi,\bmalpha]$ as discussed in Section
\ref{Section:AsymptoticExpansions}. The reminder $ o_1(\rho^N)$ is
assumed to be, at least, of class $C^1$. 
\end{assumption}

\begin{remark}
{\em Making use of a generalisation of the estimates near $\mathcal{I}$
introduced in \cite{Fri03b} for the massless spin-2 filed it is, in principle, possible to relate,
in a rigorous manner, Taylor-like expansions like the one in
Assumption \ref{Assumption:TaylorScalarField} arising from the jets
computed in Section \ref{Section:AsymptoticExpansions} and actual
solutions to the Maxwell-scalar field. The main challenge in the
present case compared to the analysis in \cite{Fri03b} is the
non-linearity of the system of equations. The discussion of this
problem, which would allow to reduce Assumption
\ref{Assumption:TaylorScalarField} to more basic hypothesis falls,
however, outside the scope of the present article. }
\end{remark}

Consistent with Assumption \ref{Assumption:TaylorScalarField} and
following the discussion of Section \ref{Section:AsymptoticExpansions},
generically, the scalar field has, near $\mathcal{I}$ the form
\[
  \phi =\varphi_\star + {\rm O} (\rho (1- \tau) {\rm log}(1- \tau)).
\]
Now, recall that the physical scalar field $\tilde{\phi}$ is related
to the unphysical one via $\tilde{\phi}=\Theta \phi$ with $\Theta=\rho
(1- \tau^2) \approx \rho (1- \tau)$ near $\tau=1$ ---i.e
$\mathscr{I}^+$. Accordingly, one has that
\[
\tilde{\phi}= \rho (1- \tau) \varphi_\star+ {\rm O}(\rho^2
(1- \tau)^{2} {\rm ln}(1- \tau)).
\]
Finally, expressing the latter in terms of the physical radial Bondi
coordinate $\tilde{r}\approx 1-\tau$ one concludes that
\[
\tilde{\phi}= \frac{\varphi_\star}{\tilde{r}}+ O\left(\frac{\ln \tilde{r}}{\tilde{r}^2}\right).
\]
Thus, to leading order, the physical scalar field satisfies the
classic peeling behaviour. Polyhomogeneous (i.e. logarithmic
contributions) are subleading. 

\subsection{The Maxwell field}
In analogy to the discussion of the scalar field, we make the
following assumption on the components of the Maxwell spinor
---cfr. Assumption \ref{Assumption:TaylorScalarField}:

\begin{assumption}
\label{Assumption:MaxwellScalarField}
On $\mathcal{M}$, the components of the Maxwell spinor $\phi-{AB}$ satisfy the asymptotic
expansion
\[
\phi_i = \sum_{p=0}^N \frac{1}{p!}\phi_i^{(p)}\rho^p + o_1(\rho^N),
\qquad i=0,\;1,\;2,
\]
for some sufficiently large $N$ and 
where $\phi_i^{(p)}$ the coefficients contained in the jet
$J^p[\bmphi]$ of the Maxwell field which can be computed from the
solution jet
$J^{(p)}[\phi,\bmalpha]$ as discussed in Section
\ref{Section:AsymptoticExpansions}. The reminder $ o_1(\rho^N)$ is
assumed to be, at least, of class $C^1$. 
\end{assumption}

A careful inspection of the solutions to the Maxwell-scalar field
equations at order $p=1$ following the discussion in Section
\ref{Section:AsymptoticExpansions} shows that, for generic data, 
close to null infinity, $\mathscr{I}^+$, one has that 
\begin{eqnarray*}
&& \phi_0 = O\big( (1-\tau)^2\ln(1-\tau) \big), \\
&& \phi_1 =O\big( (1-\tau) \ln(1-\tau)  \big), \\
&& \phi_2 = O\big( \ln(1-\tau)  \big). 
\end{eqnarray*}
The above expressions are given in the F-gauge. To analyse the peeling
properties of solutions with this behaviour we transform into the NP
gauge making use of Proposition \ref{Proposition:ChangeOfGauge}. More
precisely, the physical components of the Maxwell spinor in the NP
gauge $\tilde{\phi}_0$, $\tilde{\phi}_1$, $\tilde{\phi}_2$, are given
by:
\begin{eqnarray*}
&&\tilde{\phi}_0 =\Theta \Lambda{}^P{}_0 \Lambda{}^Q{}_0 \phi_{PQ}, \\
&&\tilde{\phi}_1 =\Theta \Lambda{}^P{}_1 \Lambda{}^Q{}_0 \phi_{PQ},\\
&&\tilde{\phi}_2 = \Theta\Lambda{}^P{}_1 \Lambda{}^Q{}_1 \phi_{PQ}.
\end{eqnarray*}
Observing that, to leading order, the physical Bondi radial coordinate satisfies
$\tilde{r}\approx 1-\tau$, one concludes that 
\[
  \tilde{\phi}_0 =O\left(  \frac{\ln \tilde{r}}{\tilde{r}^3}
   \right),
  \qquad \tilde{\phi}_1 = O\left(  \frac{\ln \tilde{r}}{\tilde{r}^2}
   \right), \qquad  \tilde{\phi}_2 = O\left( \frac{\ln \tilde{r}}{\tilde{r}}  \right).
 \]
 The key point to notice in the above expressions is the presence of a
 logarithm in the leading term of the radiation field ---$\phi_2$ in the
 conventions used in this article. This
 is a  specific property of the Maxwell-scalar field ---in a decoupled
 text Maxwell field on flat spacetime the behaviour of this particular
 component is always
 \[
\phi_2=O\left( \frac{1}{\tilde{r}} \right),
\]
see, e.g. \cite{Val00a}.

\section{Conclusions}
\label{Conclusions}
The study of the non-linear interaction between a Maxwell and a scalar
field shows a more singular behaviour than what can be expected by
studying the behaviour of the fields when non interacting. The cubic coupling
in the Maxwell-scalar field equations generically makes the solutions
more singular than what it would be expected from the mere analysis of
the linear analogue. This situation stands in stark contrast to that
of systems with quadratic coupling like that of the Einstein field
equations for which the solutions to the homogeneous transport
equations in both the linear and full non-linear case share
the same type of logarithmic divergences. In this sense, the
Maxwell-scalar field is not a good toy model to analyse the effects of
non-linear interactions in a neighbourhood of spatial infinity. A
model which potentially overcomes this shortcoming is the
Dirac-Maxwell system for which the coupling is quadratic ---this will
be discussed elsewhere.

Finally, we observe that for generic initial data which have finite energy and are
analytic around $\mathcal{I}$ the solution to the transport equations
on $\mathcal{I}$ have logarithmic singularities at the critical sets
$\mathcal{I}^+$ and $\mathcal{I}^-$. The propagation of the singularities at $\mathcal{I}^\pm$ 
along the conformal boundary will destroy the smoothness of the Faraday
tensor and the scalar field tensor at $\mathcal{I}^\pm$ so that, in contrast to a decoupled
context, there is no peeling behaviour.

\section*{Acknowledgements}
We have made substantial use of the suite {\tt xAct} for the {\tt
  Wolfram programming language} ---see \cite{xAct}.


\appendix

\section{The tracefree Ricci spinor}
\label{Appendix:RicciTensor}
The use of commutators to obtain the wave equations satisfied by the
components of the gauge potential leads to terms involving the
spinorial counterpart of the tracefree Ricci spinor ---see equation \eqref{MSFWave2}.  The symmetries
of the tracefree Ricci spinor $\Phi_{AA'BB'}$ imply the decomposition
\begin{eqnarray*}
&& \Phi_{AB'CD'} = \tfrac{1}{6}\Phi_{AC}\epsilon_{B'D'}
   +\tfrac{1}{3}\Phi_{CB}\tau_{AD'}\tau^B{}_{B'}
   +\tfrac{1}{6}\Phi_{CB}\tau_{AB'}\tau^B{}_{D'}\\
 &&
    \hspace{2cm} +\tfrac{1}{6}\Phi_{AB}\tau^B{}_{D'}\tau_{CB'}
   +\tfrac{1}{3}\Phi_{AB}\tau^B{}_{B'}\tau_{CD'}
    +\Phi_{ABCD}\tau^B{}_{B'}\tau^D{}_{D'} \\
&& \hspace{2cm}+\tfrac{1}{3}\Phi h_{ACBD}\tau^B{}_{B'}\tau^D{}_{D'},
\end{eqnarray*}
where 
\[
\Phi_{AB}=\Phi_{(AB)}, \qquad \Phi_{ABCD}=\Phi_{(ABCD)}. 
\]
A direct computation of the components of the Schouten tensor of the Weyl connection associated to the covector
$f_{AA'}$ in the conformal
representation of the Minkowski spacetime given in the F-gauge shows
that all its components vanish. Observing that 
\[
\hat{L}_{ba} = L_{ab} +f_a f_b -\frac{1}{2}f_c f^c g_{ab} -\nabla_b f_a,
\]
it follows then that 
\[
\Phi_{ABA'B'} = f_{AA'} f_{BB'} -\frac{1}{2}f_{CC'} f^{CC'}
\epsilon_{AB}\epsilon_{A'B'} -\nabla_{BB'}f_{AA'}.
\]
In the present case one has that
\[
f_{AA'} =-x_{AB} \tau^B{}_{A'},
\]
consistent with the fact that $f_{AA'}\tau^{AA'}=0$. Combining the
above observations one can conclude that
\[
\Phi=-1, \qquad \Phi_{AB} =0, \qquad \Phi_{ABCD}= x_{(AB} x_{CD)}. 
\]

\section{Properties of the solutions to the Jacobi ordinary
  differential equation}
\label{Appendix:JacobiODE}

In the following it will be convenient to define
\begin{equation}
D_{(n,\alpha,\beta)} a \equiv (1-\tau^2) \ddot{a}
+\big(\beta-\alpha-(\alpha+\beta+2)\tau  \big) \dot{a} +n(n+\alpha+\beta+1)a,
\label{JacobiOperator}
\end{equation}
so that the general \emph{Jacobi equation} can be written as
\begin{equation}
D_{(n,\alpha,\beta)} a=0.
\label{JacobiEquationGeneric}
\end{equation}
A class of solutions to  \eqref{JacobiEquationGeneric} is given by the
\emph{Jacobi polynomial} of degree $n$ with integer parameters
$(\alpha,\beta)$ given by
\[
P_n^{(\alpha,\beta)}(\tau) \equiv \sum_{s=0}^n
\binom{n+\alpha}{s}\binom{n+\beta}{n-s}\left( \frac{\tau-1}{2}\right)^{n-s}\left(\frac{\tau+1}{2}\right)^s.
\]
It follows from the above that 
\[
P_0^{(\alpha,\beta)}(\tau)=1,
\]
and that
\[
P_n^{(\alpha,\beta)}(-\tau) =(-1)^n P_n^{(\beta,\alpha)}(\tau).
\]
Solutions to \eqref{JacobiEquationGeneric} satisfy the identities
\begin{subequations}
\begin{eqnarray}
&&\hspace{-1.5cm} D_{(n,\alpha,\beta)} \left( \left( \frac{1-\tau}{2}
   \right)^{-\alpha}a(\tau)  \right) = \left( \frac{1-\tau}{2}
   \right)^{-\alpha} D_{(n+\alpha,-\alpha,\beta)} a(\tau), \label{JacobiEqnIdentity1}\\
&&\hspace{-1.5cm} D_{(n,\alpha,\beta)} \left( \left( \frac{1+\tau}{2}
    \right)^{-\beta}a(\tau)  \right) = \left( \frac{1+\tau}{2}
    \right)^{-\beta} D_{(n+\beta,\alpha,-\beta)} a(\tau), \label{JacobiEqnIdentity2}\\
&&\hspace{-1.5cm} D_{(n,\alpha,\beta)} \left( \left( \frac{1-\tau}{2}
   \right)^{-\alpha}\left( \frac{1+\tau}{2}
    \right)^{-\beta}a(\tau)  \right) = \left( \frac{1-\tau}{2}
   \right)^{-\alpha}\left( \frac{1+\tau}{2}
    \right)^{-\beta} D_{(n+\alpha+\beta,-\alpha,-\beta)} a(\tau), \label{JacobiEqnIdentity3}
\end{eqnarray}
\end{subequations}
which hold for $|\tau|<1$, arbitrary $C^2$-functions $a(\tau)$ and
arbitrary values of the parameters $\alpha$, $\beta$, $n$.

\medskip
An alternative definition of the Jacobi polynomials, convenient for
verifying when the functions vanish identically, is given by
\[
P_n^{(\alpha,\beta)}(\tau) = \frac{1}{n!}\sum_{k=0}^n c_k \left( \frac{\tau-1}{2} \right)^k
\]
with
\begin{eqnarray*}
&& \hspace{-1cm}c_0\equiv (\alpha+1)(\alpha+2)\cdots(\alpha +n), \\
&& \hspace{2cm}\vdots \\
&& \hspace{-1cm} c_k \equiv \frac{n!}{k!(n-k)!}(\alpha+k+1)(\alpha+k+2)\cdots
   (\alpha+n)\times (n+1 +\alpha +\beta) (n+2+\alpha+\beta)\cdots
   (n+k+\alpha+\beta),\\
&& \hspace{2cm}\vdots \\
&& \hspace{-1cm}c_n\equiv (n+1+\alpha+\beta)(n+2+\alpha+\beta) \cdots (2n+\alpha+\beta).
\end{eqnarray*}
Thus, for example, for $\alpha=\beta=-p$ and $n=p+\ell$ one finds that
the string of products in the above coefficients start at a negative
integer value and end up at a positive one indicating that one of the
factors vanishes. Accordingly, the whole coefficient must vanish.

\section{Details on the computation of the series solutions}
\label{Appendix:SeriesSolution}
The purpose of this appendix is to discuss some of the details in the
computation of the series solutions presented in Proposition
\ref{Proposition:BehaviourHomogeneousEqnsCoupledCase}. The approach
followed here is a variation of the classical Frobenius method ---see
e.g. \cite{Tes11}, Chapter 4. 

\subsection{The first solution}
Following the main text, we consider the model equation 
\begin{equation}
 (1- \tau^2)\ddot{\zeta}{} + 2(\varsigma + (n-1)\tau)\dot{\zeta}+
 (\ell(\ell+1)-n(n-1)-\varkappa){\zeta}=0,
 \label{ModelEquation:Appendix}
\end{equation}
and we look for solutions satisfying the Ansatz
\begin{equation}
  {\zeta}= (1- \tau)^r \sum_{k=0}^{\infty} D_k (1- \tau)^k, \qquad D_0
  \neq 0.
  \label{Ansatz:ModelEqnAppendix}
\end{equation}
Differentiation of this power series gives 
\begin{eqnarray*}
&&\dot{\zeta}=(-1)\sum_{k=0}^{\infty}(k+r) D_k (1- \tau)^{k+r-1}, \\
&&\ddot{\zeta}=\sum_{k=0}^{\infty}(k+r) (k+r-1) D_k (1- \tau)^{k+r-2}.
\end{eqnarray*}
Hence,  observing that 
\[
  2 (n-1)\tau =  2 (n-1)(1-(1-\tau))
\]
and by replacing the derivatives into the model equation
\eqref{ModelEquation:Appendix} one obtains 
the indicial polynomial
\begin{equation}
  (2r(r-1)-2\varsigma r-2(n-1)r)=0.
  \label{IndicialPolynomial}
\end{equation}
Accordingly, one has that for $\varsigma=-1$ one has the roots $r_1=0$
and $r_2=n-1$; for $\varsigma=1$ the roots are $r_1=0$ and $r_2=n+1$;
whereas for $\varsigma=0$ one has the roots $r_1=0$ and $r_2=n$. Some
further lengthy manipulations lead to the following recurrence
relations for the coefficients in the Ansatz
\eqref{Ansatz:ModelEqnAppendix}:
\begin{itemize}
\item[(i)] for $\varsigma=1$ and $r_1=0$
\[
 D_{k+1} =-\frac{k (k-3)+2n + (n(1-n)+\ell(\ell+1)-\varkappa)}{2k (k-n-1)} D_k,
\]
while if $r_2=n+1$ one has
\[
D_{k+1} = - \frac{(k+n+1) (k+n)+ (n(1-n)+\ell(\ell+1)-\varkappa) -2(k+1)}{2(k+n+1) (k+n-1) } D_k;
\]

\item[(ii)] for $\varsigma=-1$ and $r_1=0$ one has 
\[
D_{k+1}=-\frac{k (k-3) + 2n + (n(1-n)+\ell(\ell+1)-\varkappa)}{2k(k-n+1)}  D_k,
\]
while if $r_2=n-1$ one has
\[
D_{k+1}=- \frac{(k+n-1) (k+n-4) + 2n  + (n(1-n)+l(l+1)-x)}{2k(k+n-1) }  D_k;
\]

\item[(iii)] finally, if $\varkappa=0$ and  $r_1=0$ one has that 
\[
 D_{k+1}=- \frac{k (k-3) + 2n  + (n(1-n)+\ell(\ell+1)-\varkappa)}{2k (k-n+1)}  D_k, 
\]
while if $r_2=n$ one has 
\[
D_{k+1}=- \frac{(k+n) (k+n-3) +2n + (n(1-n)+\ell(\ell+1)-\varkappa)}{2k(k+n)
}D_k.
\]

\end{itemize}

\medskip
Two key observations can be drawn from the previous expressions:
\begin{itemize}
\item[(i)] all the recurrence relations associated to non-zero roots
of the indicial polynomial are well defined for $k\geq
0$. Accordingly, these lead to an infinite Taylor series for the
solutions. These series can be resumed as hypergeometric
functions. These solutions are regular and, in fact, analytic at
$\tau=1$. Analogous series solutions can be obtained for $\tau=-1$. 

\item[(ii)] All the recurrence relations associated to zero roots of
  the indicial polynomial become singular for a certain non-zero value
  of $k$. Accordingly, these recurrence relations do not lead to well
  defined series solutions. 
\end{itemize}

In summary, the procedure described in this section only provides one
independent solution to the second order model ordinary differential
equation \eqref{ModelEquation:Appendix}. 

\subsection{The second solution}
Motivated by the \emph{method of reduction of order} we look for a
second solution to the model equation \eqref{ModelEquation:Appendix}
of the form
\begin{equation}
  {\zeta}=\sum_{k=0}^{\infty} G_k (1- \tau)^{k}+\ln(1- \tau) \sum_{k=0}^{\infty} M_k (1- \tau)^{k+r}, \qquad
  G_0 \neq 0, \qquad  M_0 \neq 0.
  \label{AnsatzLogarithm}
\end{equation}
The substitution of the Ansatz \eqref{AnsatzLogarithm} into equation
\eqref{ModelEquation:Appendix} leads again to the indicial polynomial
\eqref{IndicialPolynomial}. Moreover, the coefficients $M_k$ can be
shown to satisfy, for the various choices of the parameter
$\varsigma$, the same recurrence relations as in the previous
section. Accordingly, \emph{in the following we only consider the non-zero
  roots of the indicial polynomial} and the series
\[
\sum_{k=0}^{\infty} M_k (1- \tau)^{k+r_2}
\]
is a formal solution to the model equation
\eqref{ModelEquation:Appendix}. The rest of the analysis is split in
cases corresponding to the possible values of $\varsigma$.

\medskip
\noindent
\textbf{The case $\varsigma=0$.} In this case the root of the indicial
polynomial is $r=n$. For $k\leq n-2$ one has the recurrence relation
\[
G_{k+1} =\frac{ k(k-1) -2(n-1)k - (n(1-n)+\ell(\ell+1)) }{2k(k+1) -2(n-1)(k+1) } G_k.
\]
For $k= n-1$ one has
\[
  G_{n-1} =\frac{-2n}{ -(n-1)(n-2) +2n(n-1)  -2(n-1) + (n(1-n)+\ell(\ell+1)) }M_{0}.
\]
For $0\geq k\geq n$ one has
\[
  \begin{split}
G_{k+1}=& \frac{k(k-1) -2nk +2k - (n(1-n)+\ell(\ell+1)) }{2k(k+1) -2(n-1)(k+1)}G_k + \frac{-2k-2(k+1)+2(n-1)}{2k(k+1) -2(n-1)(k+1)} M_{k-n+1}\\
&+\frac{(k-1) +k-2n +2}{2k(k+1) -2(n-1)(k+1)}M_{k-n}.
\end{split}
\]
In conclusion, in the case $\varsigma=0$ one obtains a second linearly
independent solution which contains a logarithmic singularity at
$\tau=1$. This solution has only one undetermined constant ---namely $M_0$.

\medskip
\noindent
\textbf{The case $\varsigma=1$.} In this case the non-zero root of the indicial
polynomial is given by  $r_2=n+1$. For $k\leq n-1$ we have the recurrence relation
\[
 G_{k+1} = \frac{k(k-1)-2(n-1)k - (n(1-n)+\ell(\ell+1)-\varkappa)}{2k(k+1)- 2(k+1) -2(n-1)(k+1)} G_k.
\]
For $k=n$ we have the recurrence relation
\[
G_n=-\frac{(n+2)}{ -n(n-1) +2n(n-1)+ (n(1-n)+\ell(\ell+1)-\varkappa)}M_{0}. 
\]
For $k\geq n+1$ we have the recurrence relation
\[
\begin{split}
 G_{k+1} = &\frac{k(k-1)  -2(n -1)k -(n(1-n)+\ell(\ell+1)-\varkappa)}{2k(k+1) - 2(k+1)  -2(n-1)(k+1)} G_k+ \frac{-2k-2(k+1) + k +2+2(n-1)}{2k(k+1) - 2(k+1)  -2(n-1)(k+1)}M_{k-n}  \\
&+\frac{k -2n +2}{2k(k+1) - 2(k+1)  -2(n-1)(k+1)}M_{k-n-1}.
\end{split}
\]
Again, this solution has a logarithmic singularity at $\tau=1$ and the
free constant is $M_0$.

\medskip
\noindent
\textbf{The case $\varsigma=-1$.} In this case the non-zero root of
the indicial polynomial is given by $r_2=n-1$. If $k\leq n-3$, we have the recurrence relation
\[
G_{k+1} = \frac{k(k-1)-2(n-1)k - (n(1-n)+\ell(\ell+1)-\varkappa) }{2(k+1)(k-n+2)}G_k.
\]
If $k=n-2$ we have
\[
G_{n-2} =-\frac{2 (n-1)}{(n-2)(n+1) + (n(1-n)+\ell(\ell+1)-x) } M_{0}.
\]
If $k\geq n-1$ the recurrence relation is the following 
\[
  \begin{split}
 G_{k+1}  =&\frac{k(k-1)  -2(n-1)k  - (n(1-n)+\ell(\ell+1)-\varkappa)}{2(k+1) (k-n+2)} G_k -\frac{ 4(k+1) -2(n-1)}{2(k+1) (k-n+2)}M_{k-n+2}  \\
&  + \frac{2k -2n +1}{2(k+1) (k-n+2)}M_{k-n+1}.
\end{split}
\]

\medskip
All the above expressions lead to well-defined formal series solutions
to the model equation \eqref{ModelEquation:Appendix} containing a
logarithmic singularity at $\tau=1$. The regularity of the solutions
is regulated by the value of the corresponding root to the indicial
polynomial. For example, for $\varsigma=0$, the logarithmic part of
the solution contains the factor
\[
\ln(1-\tau) (1-\tau)^n.
\]
Accordingly, the first $n$ derivatives of the solution are finite at
$\tau=1$.

\medskip
The analysis sketched in this appendix is summarised in Proposition
\ref{Proposition:BehaviourHomogeneousEqnsCoupledCase} in the main
text.

\section{Solving the inhomogeneous transport equations}
\label{Appendix:VariationParameters}

In this section we discuss a general procedure to compute the
solutions to the inhomogeneous equation 
\[
(1- \tau^2)\ddot{\zeta} + 2(\varsigma + (n-1)\tau)\dot{\zeta}+
(n(1-n)+\ell(\ell+1)-\varkappa){\zeta}= f(\tau), \qquad \varsigma=-1,\,0,\,1.
\]
In the following, for convenience we write the latter in the form
\begin{equation}
\ddot{\zeta} +\frac{ 2(\varsigma + (n-1)\tau)}{ (1- \tau^2)}\dot{\zeta}+
\frac{(n(1-n)+\ell(\ell+1)-\varkappa)}{ (1- \tau^2)}{\zeta}=
\tilde{f}(\tau), \qquad \tilde{f}(\tau) \equiv
\frac{f(\tau)}{1-\tau^2}.
\label{JacobiEqnInhomogeneous}
\end{equation}

\medskip
Let, in the following $\zeta_1$ and $\zeta_2$ denote solutions to the homogeneous problem
\[
\ddot{\zeta} +\frac{ 2(\varsigma + (n-1)\tau)}{ (1- \tau^2)}\dot{\zeta}+ \frac{(n(1-n)+\ell(\ell+1)-\varkappa)}{ (1- \tau^2)}{\zeta}=0.
\]
We follow the method of variation of the parameters and look for solutions of the form
\[
{\zeta}(\tau) =A_1(\tau){\zeta}_1(\tau)+ A_2(\tau){\zeta}_2(\tau)
\]
subject to the restriction
\[
\dot{A}_1\zeta_1 + \dot{A}_2 \zeta_2 =0.
\]
A calculation readily yields
\begin{eqnarray*}
&& \dot{\zeta} = A_1 \dot{\zeta}_1 + A_2\dot{\zeta}_2, \\
&& \ddot{\zeta} = A_1\ddot{\zeta}_1 + A_2\ddot{\zeta}_2 +\dot{A}_1\dot{\zeta}_1 +\dot{A}_2\dot{\zeta}_2,
\end{eqnarray*}
so that by replacing these relations into \eqref{JacobiEqnInhomogeneous} one has that
\[
\dot{A}_1 \dot{\zeta}_1 + \dot{A}_2\dot{\zeta}_2 =\tilde{f}.
\]
Accordingly, one obtains the algebraic system
\begin{eqnarray*}
&& \zeta_1 \dot{A}_1 + \zeta_2\dot{A}_2 =0, \\
&& \dot{\zeta}_1 \dot{A}_1 + \dot{\zeta}_2\dot{A}_2 =\tilde{f}.
\end{eqnarray*}
For convenience, we rewrite this in matricial form as
\begin{equation}
\left( 
\begin{array}{cc}
\zeta_1 & \zeta_2 \\
\dot{\zeta}_1 & \dot{\zeta}_2
\end{array}
\right) 
\left(
\begin{array}{c}
\dot{A}_1\\
\dot{A}_2
\end{array}
\right)=
\left(
\begin{array}{c}
0 \\
\tilde{f}
\end{array}
\right). \label{MatricialSystem}
\end{equation}
The latter can be recast as 
\[
\left(
\begin{array}{c}
\dot{A}_1\\
\dot{A}_2
\end{array}
\right)= \frac{1}{ (1-\tau^2)W(\tau)}\left( 
\begin{array}{cc}
\dot{\zeta}_2 & -\zeta_2 \\
-\dot{\zeta}_1 & \zeta_1
\end{array}
\right) 
\left(
\begin{array}{c}
0 \\
f
\end{array}
\right),
\]
where 
\begin{equation}
W(\tau)\equiv \left| 
\begin{array}{cc}
\zeta_1 & \zeta_2 \\
\dot{\zeta}_1 & \dot{\zeta}_2
\end{array}
\right| = \zeta_1 \dot{\zeta}_2 - \zeta_2\dot{\zeta}_1,
\label{Definition:Wronskian}
\end{equation}
denotes the \emph{Wronskian} of the system \eqref{MatricialSystem}. It readily follows then that
\begin{eqnarray*}
&& \dot{A}_1(\tau) = -\frac{\zeta_2(\tau) f(\tau)}{(1-\tau^2)W(\tau)},\\
&& \dot{A}_2(\tau) = \frac{\zeta_1(\tau) f(\tau)}{(1-\tau^2)W(\tau)}.
\end{eqnarray*}
Integrating, we conclude that 
\begin{subequations}
\begin{eqnarray}
&& A_1(\tau) = A_{1\star} -\int_0^\tau \frac{\zeta_2(s)
   f(s)}{(1-s^2)W(s)}\mathrm{d}s, \label{VariationParameter1} \\
&& A_2(\tau) = A_{2\star} +\int_0^\tau \frac{\zeta_1(s)
   f(s)}{(1-s^2)W(s)}\mathrm{d}s, \label{VariationParameter2}
\end{eqnarray}
\end{subequations}
with $A_{1\star}$ and $A_{2\star}$ constants. 

\subsubsection*{The Wronskian}
Differentiating the definition of the Wronskian $W(\tau)$, equation
\eqref{Definition:Wronskian}, and using equation
\eqref{JacobiEqnInhomogeneous} one readily finds that
\[
  \dot{W}(\tau)=\alpha(\tau)W(\tau), \qquad  \alpha(\tau)\equiv -\frac{ 2(\varsigma + (n-1)\tau)}{ (1- \tau^2)}. 
\]
The solution to this ordinary differential equation is given by
\[
  W(\tau)=e^{A(\tau)}, \qquad \dot{A}(\tau)=\alpha(\tau).
\]
It follows then that
\[
  W(\tau)=W_\star\bigg{(}\frac{1-\tau}{1+\tau}\bigg{)}^{2\varsigma}(1-\tau^2)^{n-1},
  \qquad W_\star \;\;\mbox{a constant}.
\]
Substituting the latter expression in
\eqref{VariationParameter1}-\eqref{VariationParameter2} one obtains
the explicit expressions:
\begin{eqnarray*}
&& A_1(\tau) = A_{1\star}-\int_0^\tau \frac{a_2(s)
   f(s)}{W_\star(1-s^2)^{n}}\bigg{(}\frac{1+s}{1-s}\bigg{)}^{2\varsigma}\mathrm{d}s,
 \\
&& A_2(\tau) = A_{2\star}+\int_0^\tau \frac{a_1(s)
   f(s)}{W_\star(1-s^2)^{n}}\bigg{(}\frac{1+s}{1-s}\bigg{)}^{2\varsigma}\mathrm{d}s.
\end{eqnarray*}



\end{document}